\documentclass[12pt]{iopart}
\pdfoutput=1
 \usepackage{graphicx}
\usepackage{amssymb}
\usepackage{amsfonts}
\usepackage{color}
\usepackage[latin1]{inputenc}
\usepackage{float}
\usepackage{comment}

\newcommand{\lrangle}[1]{\langle{#1}\rangle}
\renewcommand{\vec}{\mathbf}
\newcommand{\Oc}[1]{\mathcal{O}({#1})}

\begin{document}
\title[Non-universal parameters, corrections and  
universality in KPZ growth]
{Non-universal parameters, corrections and  
universality in Kardar-Parisi-Zhang growth}
\author{Sidiney G. Alves, Tiago J. Oliveira, Silvio C. Ferreira}
\address{Departamento de F\'{\i}sica, Universidade Federal de Vi\c{c}osa, 
36570-000, Vi\c{c}osa, MG, Brazil}

\begin{abstract}
We present a comprehensive numerical investigation of non-universal parameters
and corrections related to interface fluctuations of models belonging to the
Kardar-Parisi-Zhang (KPZ) universality class, in $d=1+1$, for both flat and
curved geometries. We analyzed two classes of models. In the isotropic models
the non-universal parameters are uniform along the surface, whereas in the
anisotropic growth they vary. In the latter case, that produces curved surfaces,
 the statistics must be computed independently along fixed directions. The
ansatz $h=v_\infty t + (\Gamma t)^{1/3}\chi+\eta$, where $\chi$ is a Tracy-Widom
(geometry-dependent) distribution and $\eta$ is a time-independent correction,
is probed. 
{
Our numerical analysis shows that the non-universal parameter $\Gamma$
determined through the first cumulant leads to a very good accordance with the
extended KPZ ansatz for all investigated models in contrast with the estimates 
of $\Gamma$ obtained from higher order 
cumulants that indicate a violation of the generalized ansatz for some of the
studied models.  We associate the discrepancies to corrections of unknown nature,
which hampers an accurate estimation of $\Gamma$ at finite times.
}
The discrepancies in $\Gamma$ via
different approaches are relatively small but sufficient to modify the scaling
law $t^{-1/3}$ that characterize  the finite-time corrections due to $\eta$. Among
the investigated models, we have revisited an off-lattice Eden model that supposedly
disobeyed the shift in the mean scaling as $t^{-1/3}$ and showed that there is a
crossover to the expected regime. We have found model-dependent (non-universal)
corrections for cumulants of order $n\ge 2$. All investigated models are
consistent with  a further term of order $t^{-1/3}$ in the KPZ ansatz.
\end{abstract}


\section{Introduction}
\label{sec:intro}

Dynamics of self-affine surfaces is a fascinating topic of non-equilibrium
statistical physics where pattern formation, stochasticity, scale invariance and
universality are joined in a unified framework~\cite{barabasi,meakin}. The
KPZ equation~\cite{KPZ}
\begin{equation}
 \frac{\partial h}{\partial t} = \nu \nabla^{2} h + \frac{\lambda}{2} (\nabla
h)^{2} + \xi,
\label{eq:KPZ}
\end{equation}
proposed by Kardar, Parisi and Zhang in 1986, emerged as one of the most prominent
theoretical systems and established a universality class which encompasses a plenty of
models~\cite{barabasi,meakin}. This equation describes the non-conservative
evolution of an interface $h(x,t)$ subjected to a white noise $\xi$ defined by
$\langle \xi \rangle = 0$ and $\langle \xi(x,t)\xi(x',t') \rangle = D
\delta(x-x') \delta(t-t')$. The KPZ universality class goes beyond the surface
dynamics including theoretical studies in random polymers~\cite{Amir} and fluid
transport~\cite{Beijeren}. The understanding of the KPZ
universality class has undergone a noticeable progress during the last few years
with its experimental realization~\cite{TakeSano, TakeuchiSP, TakeuchiJSP12,Yunker}  as
well as the achievement of analytical solutions~\cite{Amir,SasaSpo1, Calabrese, 
SasaSpohnJsat,Imamura,Doussal} of the KPZ equation in 1+1 dimensions.

Self-affine interface evolution is featured by a dynamical regime where the interface
width $w$, defined as the standard deviation of the surface heights, 
$ w\equiv \sqrt{\left\langle h^2  \right\rangle - \left\langle h \right\rangle^2}$, increases in
time as a power law $w\sim t^\beta$, and a stationary regime where the interface
width depends on the system size as $w\sim L^\alpha$. $\beta$ and $\alpha$
are the growth and roughness exponents~\cite{barabasi}, respectively, that 
assume the exact values $\beta = 1/3$ and $\alpha = 1/2$ for KPZ class in $d=1+1$~\cite{KPZ}.
The universality in KPZ systems includes other important universal quantities
related to the height distributions (HDs)~\cite{TakeuchiSP,krugrev}.
Motivated by analytical determination of the HDs of some models of
the KPZ class, namely, the asymmetric exclusion process~\cite{johansson} and
polynuclear growth model~\cite{PraSpo1,PraSpo2},   Pr\"ahofer and Spohn~\cite{PraSpo1}
conjectured that the heights for any KPZ system are asymptotically  given by
\begin{equation}
h = v_\infty t + s_\lambda(\Gamma t)^{1/3} \chi
\label{eq:hdet}
\end{equation}
where $s_\lambda=\mbox{sgn}(\lambda)$, $v_\infty$ and $\Gamma$ are non-universal
parameters while $\chi$ is a stochastic variable with universal properties.
Moreover, the conjecture states that, in the dynamical regime, $\chi$ depends on the growth geometry and
is given by  Tracy-Widom (TW) distributions from random matrix theory~\cite{TW1}
for flat and curved growth geometries. The Gaussian orthogonal ensemble (GOE) is
expected for the former while Gaussian unitary ensemble (GUE) is expected for
the latter. Therefore, the KPZ universality class splits into subclasses
depending on the growth geometry. This conjecture has been verified by many
analytical results~\cite{Amir,SasaSpo1, Calabrese, Imamura,Doussal}, computer
simulations~\cite{SchehrEPL,Alves11, Oliveira12, TakeuchiJstat} and, most
excitingly, in experiments~\cite{TakeSano, TakeuchiSP, TakeuchiJSP12, Yunker}.

The TW distributions are conjectured for the asymptotic regime, but many
experimental~\cite{TakeSano,TakeuchiJSP12}, analytical~\cite{SasaSpo1,Ferrari} 
and numerical~\cite{Alves11,Oliveira12} works have shown the existence of 
a slow convergence with apparently universal properties. More specifically, 
the rescaled distribution $P(q)$, where
\begin{equation}
q = \frac{h-v_\infty t}{s_\lambda(\Gamma t)^{1/3}},
 \label{eq:hsca}
\end{equation}
has a shift in relation to the TW distributions that vanishes as 
\begin{equation}
\lrangle{q} - \lrangle{\chi} \sim t^{-1/3}.
\label{eq:shift}
\end{equation}
An  exception to the correction $t^{-1/3}$ was obtained in the partially asymmetric
simple exclusion process (PASEP) for the asymmetry parameter
$p_c=0.7822787862\ldots$~\cite{Ferrari}. The PASEP result suggests that
corrections different from $t^{-1/3}$ would appear in very particular
situations. However, Takeuchi~\cite{TakeuchiJstat} reported off-lattice
simulations of a radial Eden model where the first cumulant of the scaled
distributions converges to the asymptotic GUE value as $t^{-2/3}$, in contrast
with other Eden versions~\cite{Alves11}.
It is somehow intriguing that a stochastic model without control
parameters does not have the usual correction. 
Still in reference~\cite{Ferrari}, a deterministic shift $\eta$ was found  
in the  polynuclear growth model (PNG) and in the totally/partially
asymmetric exclusion processes (TASEP/PASEP), generalizing equation~(\ref{eq:hdet}) to 
\begin{equation}
 h = v_\infty t+s_\lambda(\Gamma t)^{1/3}\chi+\eta+\ldots
\label{eq:hpluscorr}
\end{equation}
The correction $t^{-1/3}$ in the first cumulant derives directly from equation~(\ref{eq:hpluscorr}).
It was also shown that higher order moments have corrections $\Oc{t^{-2/3}}$~\cite{Ferrari}  and, consequently,
all higher order cumulants of order $n\ge 2$ decays as $\Oc{t^{-2/3}}$ or faster. 
Sasamoto and Spohn found out a result
similar to equation~(\ref{eq:hpluscorr}) in their solution of the KPZ
equation~\cite{SasaSpo1,SasaSpohnJsat} with a difference that $\eta$ is
random, but still independent of $\chi$. 

Many models belonging to the KPZ class are analytically intractable with our
current knowledge and Monte Carlo simulations become the best accessible
method to probe the universality of their interface fluctuations. The analysis of flat
growth can be efficiently performed using lattice
models~\cite{Oliveira12,Kelling} whose computer implementations are commonly
simple, relatively quick and low memory demanding. Eden model~\cite{eden} and
its variations~\cite{meakin} are basic examples of radial growth belonging to
the KPZ class. However, it is well known that radial growth in lattices is
distorted by anisotropy effects~\cite{Zabolitzky,Batchelor99,Alves06,Paiva07}.
Therefore, off-lattice Eden simulations~\cite{Alves11,TakeuchiJstat,Ferreira06}
or suitable tricks using tuning parameters~\cite{Alves11,Paiva07} are needed to
investigate interface fluctuations in the entire surface. Although off-lattice
simulations of Eden growth are quite affordable in 1+1
dimensions~\cite{BJP}, the generalization to higher
dimensions is cumbersome~\cite{Alves12}.  


In this  work, we present extensive simulations of models belonging 
to the KPZ universality class to address the validity of the extended KPZ conjecture
including the correction $\eta$ given in equation~(\ref{eq:hpluscorr}).
We investigated models exhibiting either isotropic or anisotropic growth. 
In the isotropic growth the velocity is the same for all parts of the interface 
such that the entire profile can be used to perform statistics. On other hand,
in an anisotropic growth, in which the interface velocity varies along the interface,
we must analyze fixed directions independently and, consequently, 
a large ensemble of samples are required since one or a few points from
each sample are used for statistics.

We have found that the procedure to determine the model-dependent
parameter $\Gamma$ via second cumulant may be troublesome due to strong and/or
puzzling corrections to the scaling. We adopted an {alternative} method 
using the first cumulant derivative, that exhibits a monotonic convergence 
to the asymptotic value with a correction $t^{-2/3}$ in all investigated
models. This method allowed to determine shifts  in the first
cumulants always consistent with a decay $t^{-1/3}$, reinforcing the generality of
the KPZ ansatz. We also investigated the convergence of higher order cumulants 
and identified complex and non-universal behaviors.
Finally, an additional term
$t^{-1/3}$ in the KPZ ansatz was found in all models considered here.

The  paper is organized as follows. Section~\ref{sec:isoflat} presents the
analysis of flat growth models concomitantly with the description of the
numerical recipes applied in the present work. Sections~\ref{sec:isorad}  and
\ref{sec:anisorad} are devoted to isotropic and anisotropic radial growth,
respectively. Section~\ref{sec:droplet}
presents the analysis of the droplet growth. We draw our concluding remarks
in section~\ref{sec:conclu}.

\section{Isotropic flat growth}
\label{sec:isoflat}

We consider two versions of the restricted solid-on-solid (RSOS) model~\cite{kk}
and the ballistic deposition (BD) model~\cite{barabasi} on initially flat substrates.
Averages were computed using $10^4$ samples and system sizes  $L=2^{18}$
resulting in more than $2\times 10^9$ data points for the statistics.

The RSOS model is defined as a random deposition obeying the height
restriction $\Delta h = |h_{j}-h_{j+1}|\le m$.  In the BD
model, particles move normally to the substrate and permanently
attach to the first nearest neighbor contact with a previously deposited
particle. In both models, periodic boundary conditions are assumed 
and the time is increased by $\Delta t = 1/L$  for each
deposition attempt. 

\begin{figure}[b]
 \centering
 \includegraphics[width=9cm]{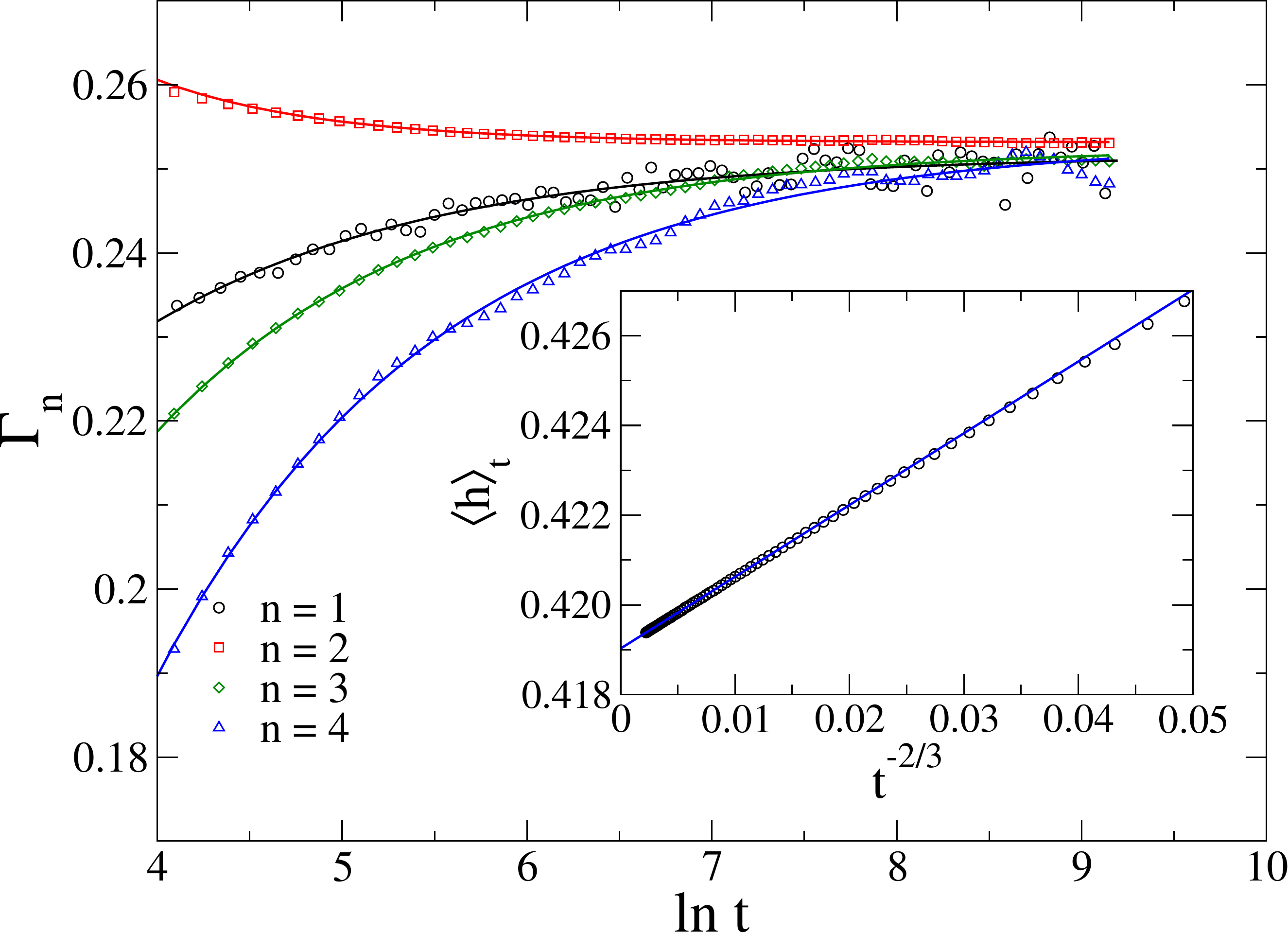}
\caption{\label{fig:rsosgamma} Numerical determination of the parameter $\Gamma$
for the RSOS model grown on flat substrates, with $\Delta h \le 1$, 
using different cumulants. Solid lines are
non-linear regressions. The inset shows the interface velocity against
$t^{-2/3}$ (symbols) and the linear regression used to determine the asymptotic
velocity.}
\end{figure}

Accurate estimates of the non-universal parameters are imperative to a
reliable characterization of the corrections. The asymptotic velocity is accurately
determined taking the time derivative of the mean height  in equation~(\ref{eq:hpluscorr}):
\begin{equation}
 \lrangle{h}_t = v_\infty+ \frac{s_\lambda\Gamma^{1/3}\lrangle{\chi}}{3} t^{-2/3}.
\label{eq:dhdt}
\end{equation}
Then, the velocity $v_\infty$ can be extracted by extrapolating $\lrangle{h}_t$
versus $t^{-2/3}$ in a linear regression for $t\rightarrow\infty$. The inset of figure~\ref{fig:rsosgamma}
shows a typical plot used to determine $v_\infty$ for RSOS model with $m=1$.
Interface velocities for all investigated isotropic models are shown in
table~\ref{tab:noniso}.

\begin{table*}[t]
\begin{center}
\begin{footnotesize}\begin{tabular}{cccccccc}
\hline\hline
model      &$s_\lambda$ &$v_\infty$  & $\Gamma_1$ & $\Gamma_2$ & $\Gamma_3$ & $\Gamma_4$  & $\lrangle{\eta}$\\ \hline
RSOS (m=1) &-1&0.419030(3) & 0.252(1)   & 0.2532(3)  & 0.2525(2)  & 0.2534(3)   & $-0.32(4)$\\ 
RSOS (m=2) &-1&0.60355(1)  & 0.812(2)   & 0.816(9)   & 0.811(2)   & 0.815(2)    & $-0.66(6)$   \\ 
BD         &1 &2.13983(1)  & 4.94(1)    & 4.778(2)   & 4.799(7)   & 4.74(2)     &$-0.9(1)$\\ 
Eden D     &1 &0.51371(2)  & 1.00(1)    & {$\approx 1$}& 0.99(1)  & 0.98(2)     &$0.50(5)$\\ \hline\hline
\end{tabular}             \end{footnotesize}
\end{center}
\caption{Non-universal parameters for distinct isotropic growth models. The number in parenthesis
represents the uncertainties obtained from the non-linear regressions. 
The parameter $\Gamma_2$ does not have uncertainty for off-lattice Eden model 
due to the lack of a monotonic convergence.} 
\label{tab:noniso}
\end{table*}

The parameter $\Gamma$, in terms of the constants of the KPZ equation, is given by 
$\Gamma=\frac{1}{2}A^2|\lambda|$ with $A=D/\nu$~\cite{krugrev,TakeuchiJSP12}.
The parameters $\lambda$ and $A$ can be determined numerically using the tilt dependence
of the growth velocity and the amplitude of a two-point correlation function, respectively~\cite{krug}.
Alternatively, $\Gamma$ can be obtained directly from equation~(\ref{eq:hdet})
assuming that the cumulants of $\chi$ are those of TW distributions. The scaled second cumulant
\begin{equation}
 g_2 = \frac{\lrangle{h^2}_c}{t^{2/3}}\rightarrow \Gamma^{2/3}\lrangle{\chi^2}_c,
\label{eq:g2}
\end{equation}
where $\lrangle{X^n}_c$ denotes  the $n$th cumulant of $X$,
has been used to this purpose~\cite{TakeSano,TakeuchiJstat,TakeuchiSP,Oliveira12}.
In principle, one can obtain $\Gamma$ from
any higher order cumulant since equation~(\ref{eq:hdet}) yields
\begin{equation}
 g_n=\frac{\lrangle{h^n}_c}{s_\lambda^n ~ t^{n/3}}\rightarrow \Gamma^{n/3}\lrangle{\chi^n}_c, ~~~~ n\ge 2
\label{eq:gn}
\end{equation}

Alternatively, the  parameter $\Gamma$ can also be obtained from the first cumulant since
\begin{equation}
g_1= s_\lambda 3(\lrangle{h}_t-v_\infty)t^{2/3}\rightarrow \Gamma^{1/3}\lrangle{\chi}.
\label{eq:g1}
\end{equation}
Although widely used in many experimental~\cite{TakeuchiJSP12,TakeSano,TakeuchiSP} 
and computer~\cite{Oliveira12,TakeuchiJstat} studies, 
the determination of $\Gamma$ using second order cumulants has complications 
when there are statistical dependencies among $\chi$, $\eta$ or other unknown terms
in equation~(\ref{eq:hpluscorr}). In this case, crossed terms  appear in cumulants
leading to relevant corrections that, in principle, may be puzzling. Otherwise,
the analysis using $\lrangle{h}_t$ is free from crossed terms and does not depend
on $\eta$. Noise in
numerical derivatives counts against this last method. We propose that the first
cumulant derivative yields more reliable estimates of $\Gamma$ than 
$\lrangle{h^2}_c$ does, as discussed in the this and in the following sections.

\begin{figure}[hb]
 \centering
 \includegraphics[width=9cm]{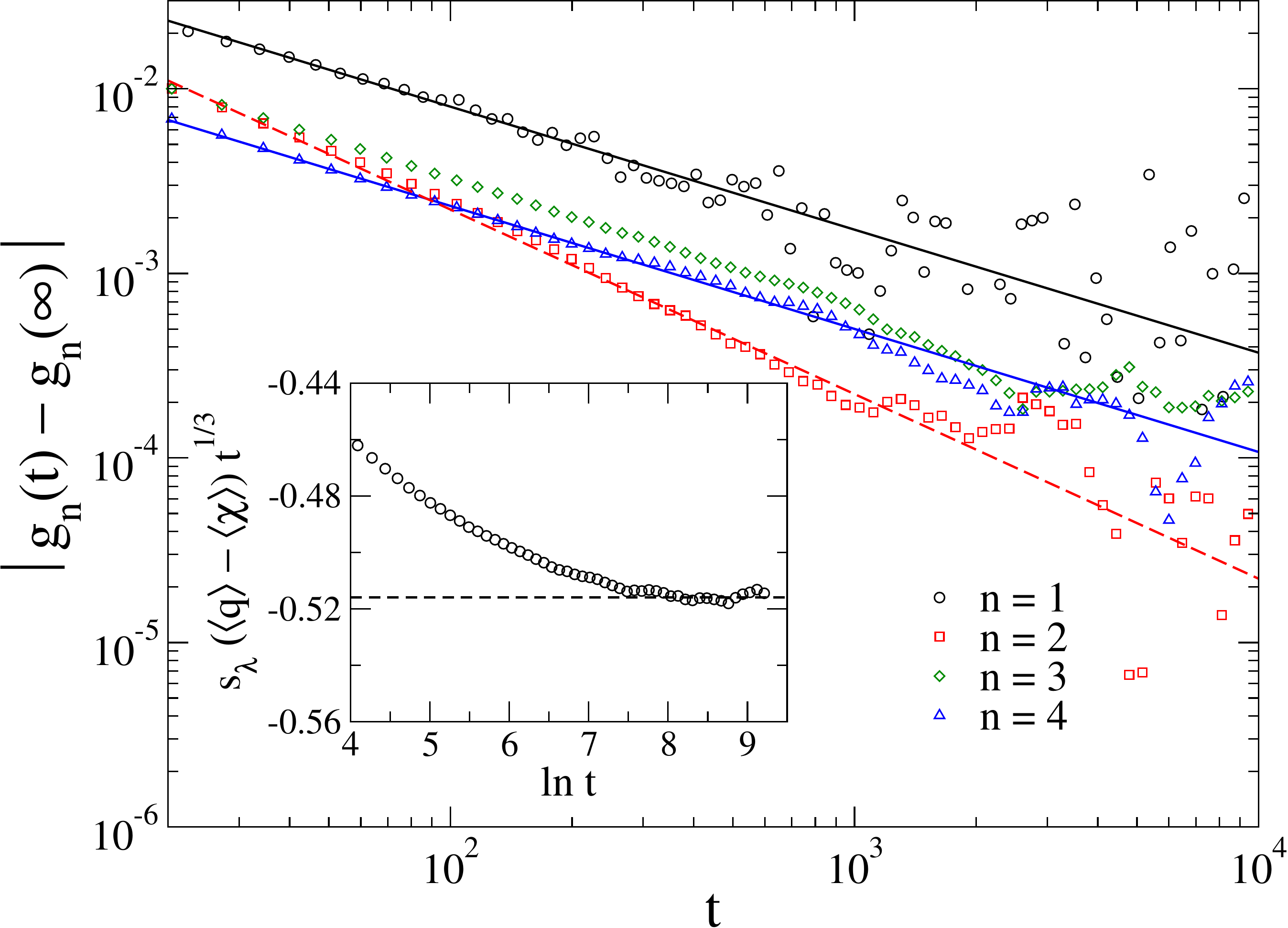}
\caption{\label{fig:rsos} Corrections in $g_n$ for the RSOS model grown on flat substrates.
Dashed line represents a power law $t^{-1}$ and the solid
ones represent the power law $t^{-2/3}$. Inset shows the difference between
the first cumulant of equation~(\ref{eq:hsca}) 
and the GOE mean $\lrangle{\chi}=-0.76007$ using $v_\infty=0.419030$
and $\Gamma = 0.252$. The shift is scaled by the theoretic power law $t^{-1/3}$.}
\end{figure}

We denote by $\Gamma_n = (g_n/\lrangle{\chi^n}_c)^{3/n}$, the value of $\Gamma$
estimated via the $n$th cumulant. Figure~\ref{fig:rsosgamma} shows the curves used to
determine the non-universal parameters of the RSOS model with $m=1$.
Independently of the used cumulant, the data have finite-time corrections. To
estimate the asymptotic value as well as the scaling of the correction, numerical
values of $\Gamma_n(t)$, that are recorded exponentially spaced in time, are
plotted as function of $s = \ln t$. Assuming that there is a power law correction,
we performed non-linear regressions in the form 
\begin{equation}
\Gamma_n=\Gamma_n(\infty)+c\exp(-a_n s),
\label{eq:extrapo}
\end{equation}
discarding very short times. The extrapolated values for $\Gamma_n$
are shown in table~\ref{tab:noniso}. 
All cumulants yield essentially the same asymptotic value for $\Gamma_n$, that
is in very good agreement with our previous estimate using longer growth times 
(and less samples) without extrapolations~\cite{Oliveira12}. The correction in
$\Gamma_1$ has an exponent  $a_1 = 0.661\approx 2/3$. Assuming that the 
next term in equation~(\ref{eq:hpluscorr}) is a power law $\zeta t^{-\gamma}$, we
have that
\begin{equation}
g_1 = \Gamma^{1/3}\lrangle{\chi}+{\lrangle{\zeta}t^{-\gamma-1/3}}.
\end{equation}
So, a correction $a_1=2/3$ in $g_1$, or equivalently in $\Gamma_1$, 
shows  that the next
leading term in equation~(\ref{eq:hpluscorr}) decays as $t^{-1/3}$. The
correction in $\Gamma_2$, that have the same exponent as
$\lrangle{q^2}_c-\lrangle{\chi^2}_c$, is $a_2 \approx 1$ while $a_3\approx a_4
\approx 2/3$ are found for the
third and forth order cumulants. These results are corroborated in
figure~\ref{fig:rsos}, where the differences between $g_n$, $n=1,..,4$, and
their asymptotic values (see table~\ref{tab:cumiso}) are plotted against time. It
is worth mentioning that corrections in cumulants were already 
studied for RSOS model~\cite{Oliveira12} and are consistent with the present
results.

The power law correction $t^{-1/3}$ in the mean is also verified with a good precision
as shown in the inset of figure~\ref{fig:rsos}, where the difference between
the scaled height and the GOE first cumulant has a plateau when rescaled by
the expected power law. Considering the variation of the plateau 
inside errors of $\Gamma$ and $v_\infty$, the estimated amplitude
of the correction is $\lrangle{\eta}\approx 0.32(4)$. The mean
$\lrangle{\eta}$ for all isotropic models are shown in table~\ref{tab:noniso}.

We also simulated the RSOS model allowing a height difference $\Delta h\le 2$.
The results are essentially the same except by a non-monotonic time-dependence
for $\Gamma_2$ that initially decays and then increases towards the asymptotic
value also with a correction faster than  $t^{-2/3}$. For other $g_n$, the
corrections are approximately $t^{-2/3}$, the same as in the $m=1$ case. The
non-universal parameters for $m=2$ and cumulant ratios are also shown in
tables~\ref{tab:noniso} and~\ref{tab:cumiso}.

\begin{figure}[b]
 \centering
 \includegraphics[width=9cm]{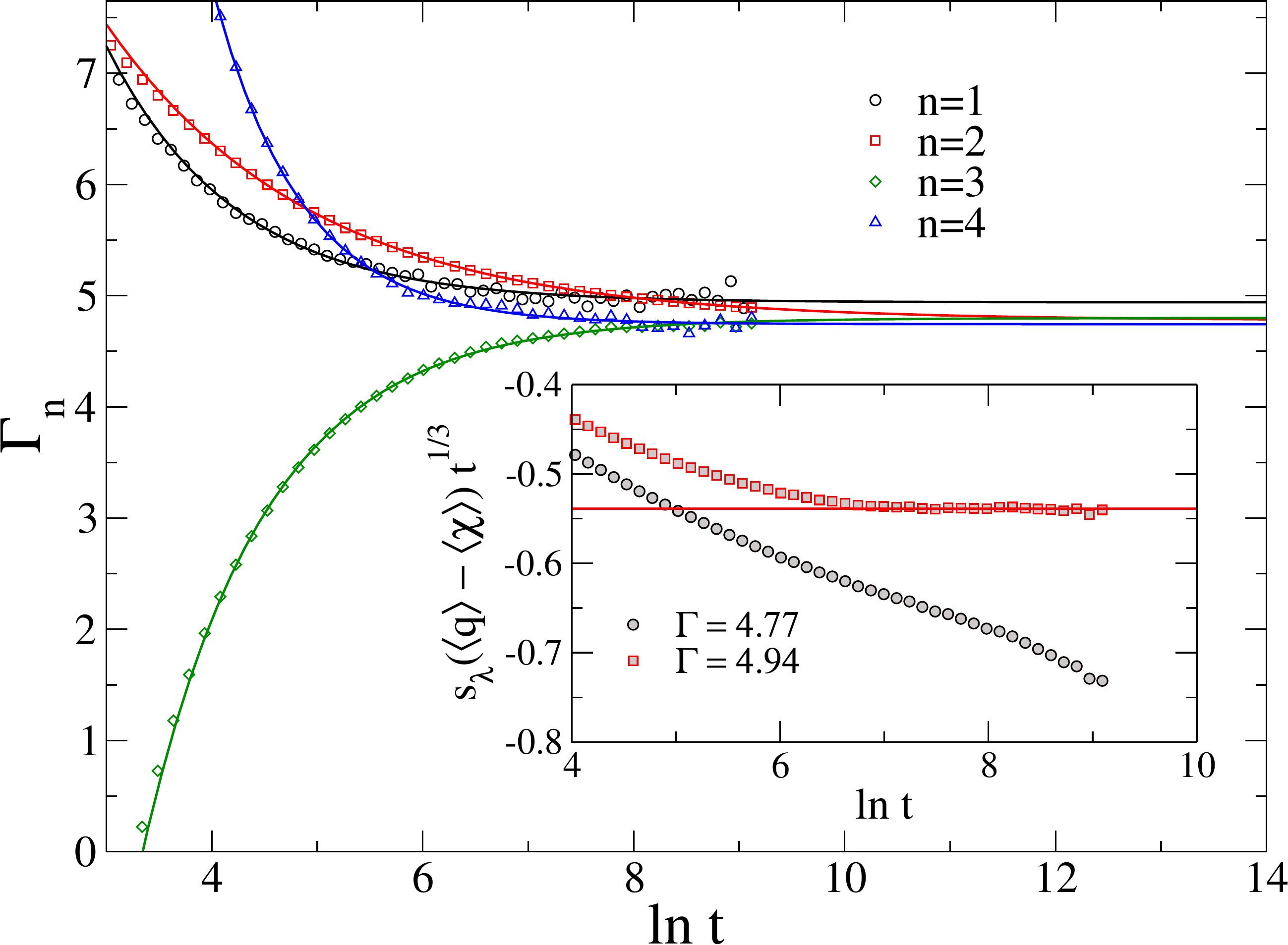}
 \caption{Determination of the non-universal parameter $\Gamma$ for the BD model
grown on flat substrates. Solid lines are non-linear regressions. Inset: Shift
in the first cumulant scaled by the expected power law $t^{-1/3}$ using two
estimates of $\Gamma$ and $v_\infty=2.13983$.}
 \label{fig:dbgamma}
\end{figure}

Figure~\ref{fig:dbgamma} shows the curves $\Gamma_n$ against time and the
respective non-linear regressions used to obtain their asymptotic values for the
BD model. Surprisingly, the extrapolation from $\Gamma_1$ is a little larger
than the others. The difference of 3-4\% is sufficient to affect the scaling law
ruling the shift of the distribution in relation to GOE as shown in the inset of
figure~\ref{fig:dbgamma}. The value $\Gamma = 4.94$ obtained from the first
cumulant derivative yields an excellent agreement with a correction  $t^{-1/3}$
while the values of $\Gamma$ extrapolated from higher order cumulants do not. 

The cause of this difference is unknown. A possibility could be crossed terms
involving random variables in higher order cumulants, that are not present in
the first one. These terms would introduce complicated corrections leading to a
non-monotonic convergence and $\Gamma$ could not be properly extrapolated using
equation~(\ref{eq:extrapo}) for $n\ge 2$. 
Another possible explanation is that the asymptotic distribution of $q$ for BD
model does not converge  to GOE but to a shifted GOE distribution with $\chi+a$
where $a$ is a deterministic and non-universal parameter. In this case, the
deterministic shift $a$ cannot be obtained using higher order cumulants since
$\lrangle{a^n}_c=0$ for $n\ge 2$. If this last hypothesis is correct, we would
have $\Gamma_1=\Gamma(1+a/\lrangle{\chi})^3$ that, according to the parameters
of table~\ref{tab:noniso}, provides a tiny shift $a\approx 0.01$. Such a small
shift is easily unnoticed in scaled height distributions obtained in simulations
or experiments. {Indeed, the correction $t^{-1/3}$ shown in the inset
of figure~\ref{fig:dbgamma} is equally obtained in curves $\lrangle{q}
-\lrangle{\chi+a}$ against $t$ (data not shown)}. It important to emphasize that
this shift was not predicted in the other KPZ systems investigated either in the
present work or elsewhere. Independently of the numerical evidence, we believe
that the non-monotonic convergence is the most likely scenario.

In reference~\cite{Oliveira12}, we have showed that the shift in the first
cumulant is consistent with the usual $t^{-1/3}$ decay  using the parameter
$\Gamma_2=4.90(5)$ obtained from $\lrangle{h^2}_c$ in simulations up to $t =
5\times 10^4$ without extrapolation. We checked that the extrapolation of the
data of reference~\cite{Oliveira12} is consistent with our current estimate of
$\Gamma_2$. A careful view in the double-logarithmic plot presented in
reference~\cite{Oliveira12} reveals a slightly bent curve indicating a
correction not so close to $t^{-1/3}$ as wished. Also, the uncertainties in
$\Gamma$ and $v_\infty$ reported there are much larger than the current ones
such that the propagated error in the power law is sufficiently large to embrace
the exponent -1/3.

\begin{figure}[h]
 \centering
 \includegraphics[width=9cm]{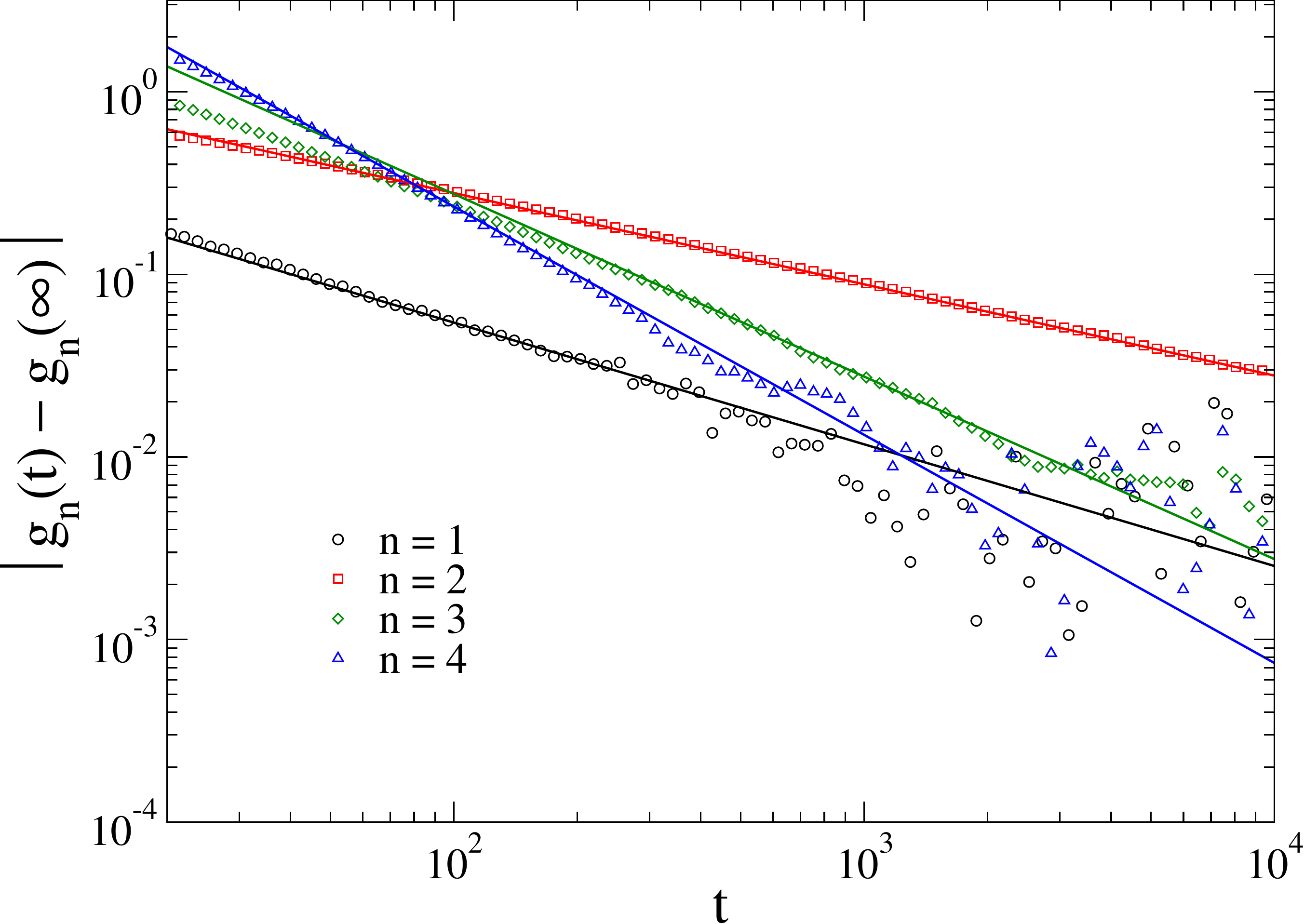}
 \caption{ \label{fig:dbcum} Corrections in $g_n$ for the  BD  model grown on
flat substrates. The lines are power laws $t^{-2/3}$ ($n=1$), $t^{-1/2}$
($n=2$), $t^{-1}$ ($n=3$), and $t^{-1.25}$ ($n=4$).}
\end{figure}

Corrections in $g_n$ for BD model are shown in figure~\ref{fig:dbcum}. In $g_1$,
we have found an exponent $a_1\approx 2/3$, the same value found for the RSOS
model. Nevertheless, an unusual exponent $a_2\approx 1/2$ was observed  in
second order cumulant. To our knowledge, there is not analytical or experimental
analogues for this correction. This exponent is at odds with our previous
analysis reporting a decay faster than $t^{-2/3}$ in plots
$\lrangle{q^2}_c-\lrangle{\chi^2}_c$ against time~\cite{Oliveira12}. The
discrepancy is due to the over-estimation of $\Gamma_2$ used in the rescaled
height $q$ while in the current analysis, the dependence on $\Gamma_2$ is
implicit to $g_2$ and, therefore, does not involve error propagation. The third
and fourth order cumulants decay with exponents $a_3\approx 0.9$ and
$a_4\approx 1.25$, respectively, that are consistent with the previous report
$\Oc{t^{-2/3}}$  or faster~\cite{Oliveira12}. These results show that the
exponents measured in the power laws related to the finite-time corrections are very
sensitive to $\Gamma$ and, thus, a precise characterization of the corrections
demands accurate estimates of $\Gamma$.
 
\section{Isotropic radial growth}
\label{sec:isorad}

\begin{figure}[b]
 \centering
 \includegraphics[width=9cm]{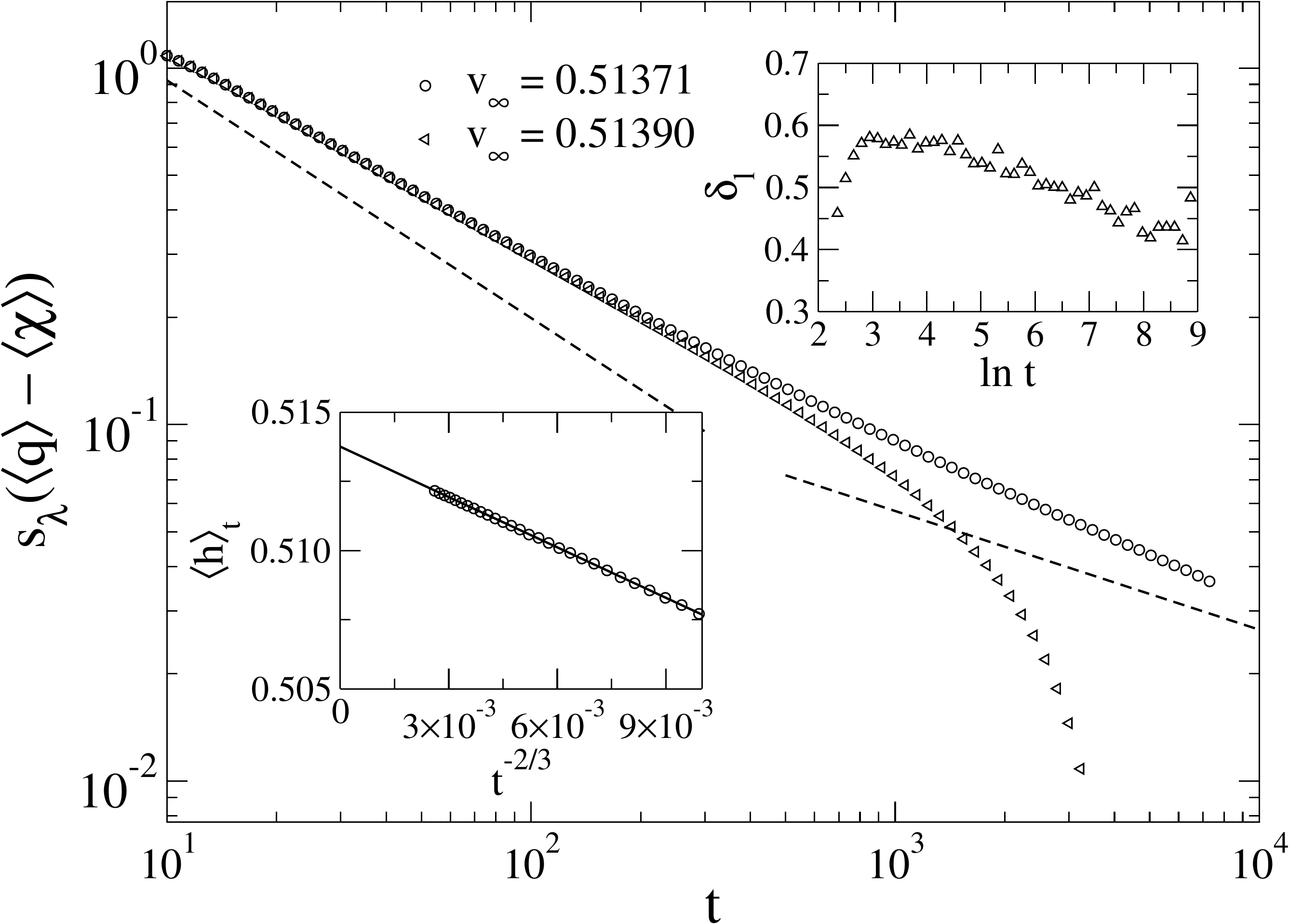}
 \caption{\label{fig:ed1st} Correction in the first cumulant of the scaled
height in relation to the corresponding GUE value
$\lrangle{\chi}=-1.77109$ for off-lattice Eden~D model. The parameters used were
$\Gamma=1.00$ and interface velocities indicated in the legends. Dashed lines
are power laws $t^{-2/3}$ and $t^{-1/3}$ as guides to the eyes. Top inset shows
the local exponent against time, while the bottom one shows the velocity against 
$t^{-2/3}$.}
\end{figure}

We simulated radial growth using an Eden model proposed in
reference~\cite{TakeuchiJstat},  that we have named as Eden D~\cite{Alves12} since
A, B, and C versions are already defined in literature~\cite{meakin,Paiva07}.
The model is defined as follows: In each time step, a particle of the cluster and 
a position tangent to it are randomly chosen. A new particle is added to the chosen 
position if  this event does not imply  the overlap with any other particle of the cluster. 
In the case of overlap, simulation proceeds to the next step. The time is increased
by $\Delta t =1/N$, where $N$ is the number of particles of the cluster at time $t$.
Optimization strategies described in reference~\cite{BJP} were used to speed up the
simulation. A cluster of diameter $8000$ takes typically 8 min of simulation in
a CPU Intel Xeon 3.2 GHz, whereas if no optimization is used the same simulation 
takes several hours of computation. 
We simulated the same version of the Eden model proposed in
reference~\cite{TakeuchiJstat} using a much better statistics (90000 samples
against 3000)  and much longer growth times ($t\simeq 8000$ against 2000).
Considering the same interval used in reference~\cite{TakeuchiJstat}
($250<t<2000$), we have found $v_\infty=0.51390$ in full agreement with
$v_\infty=0.5139(2)$ reported there. However, a more accurate estimate
$v_\infty=0.51371(1)$ was found for $t>2000$ (bottom inset of
figure~\ref{fig:ed1st}).

\begin{figure}[hbt]
 \centering
 \includegraphics[width=9cm]{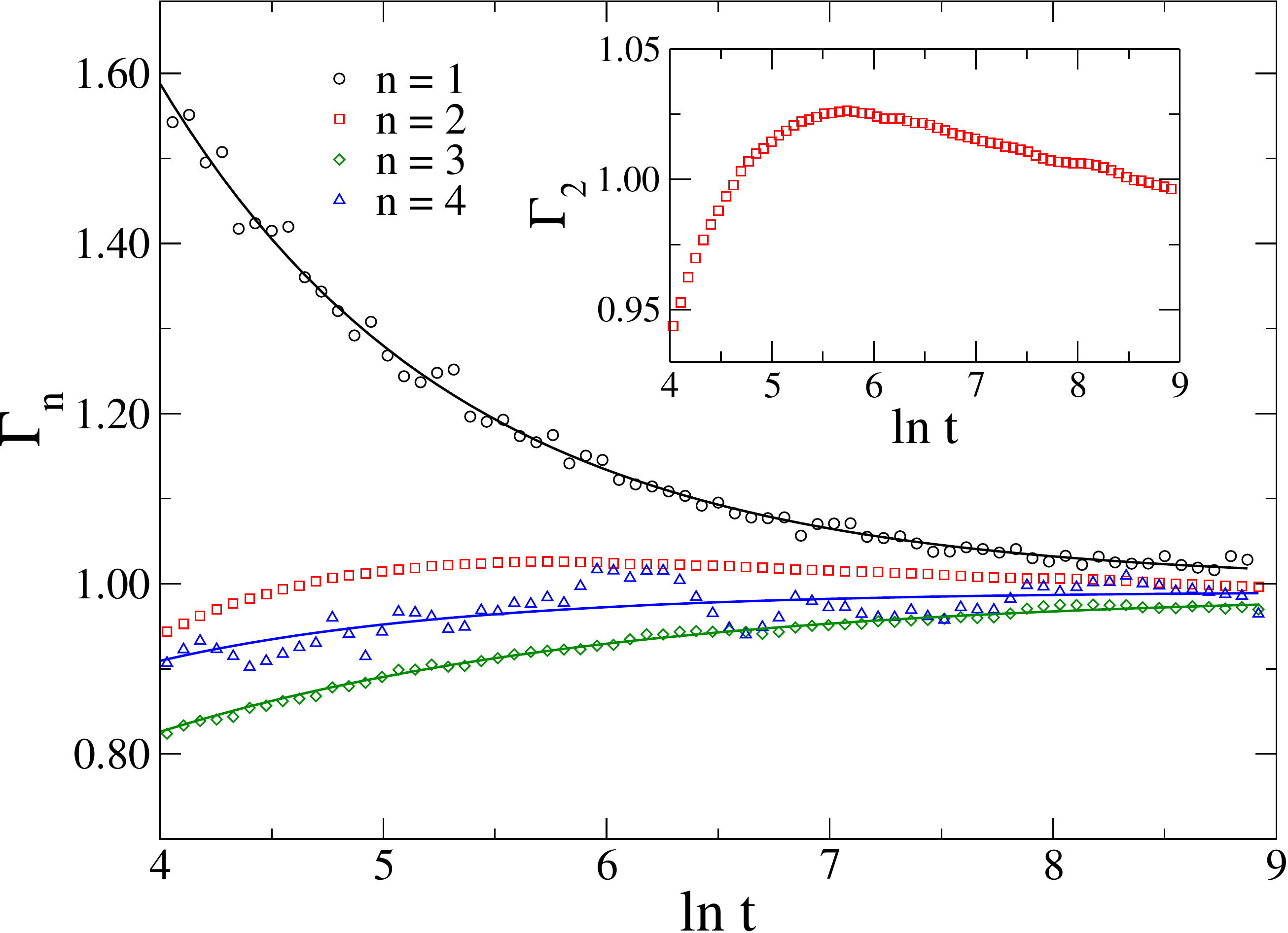}
 \caption{Determination of the non-universal parameter $\Gamma$ for the Eden model
 using different cumulants. Solid lines are non-linear
regressions used to extrapolate $\Gamma_n$. Inset: Zoom to emphasize the
non-monotonicity of $\Gamma_2$.}
 \label{fig:edgamma}
\end{figure}

Curves used to determine $\Gamma_n(\infty)$ for the Eden model are shown
in figure~\ref{fig:edgamma}. The second order cumulant depends non-monotonically
on time as highlighted in the inset of figure~\ref{fig:edgamma}. This
non-monotonicity hampers an extrapolation to $\Gamma_2(\infty)$ using equation~(\ref{eq:extrapo}).
{Using an extrapolation similar to that used in reference~\cite{TakeuchiJstat},
{\it i.e.}, forcing a correction $a_2=2/3$ in $\Gamma_2$,
we obtained $\Gamma_2\approxeq 0.995$ considering the time interval 
$1000<t<8000$. This value is slightly smaller than $\Gamma_2=1.02(2)$ reported 
in reference~\cite{TakeuchiJstat}.}
The first cumulant derivative yields a monotonic
convergence with an exponent $a_1\approx 2/3$ and extrapolates to $\Gamma_1=1.00(1)$. 
Moreover, $\Gamma_3$ and $\Gamma_4$ also have corrections consistent with $t^{-2/3}$
and converge to values very close to 1 (see table~\ref{tab:noniso}).

\begin{figure}[hbt]
 \centering
 \includegraphics[width=9cm]{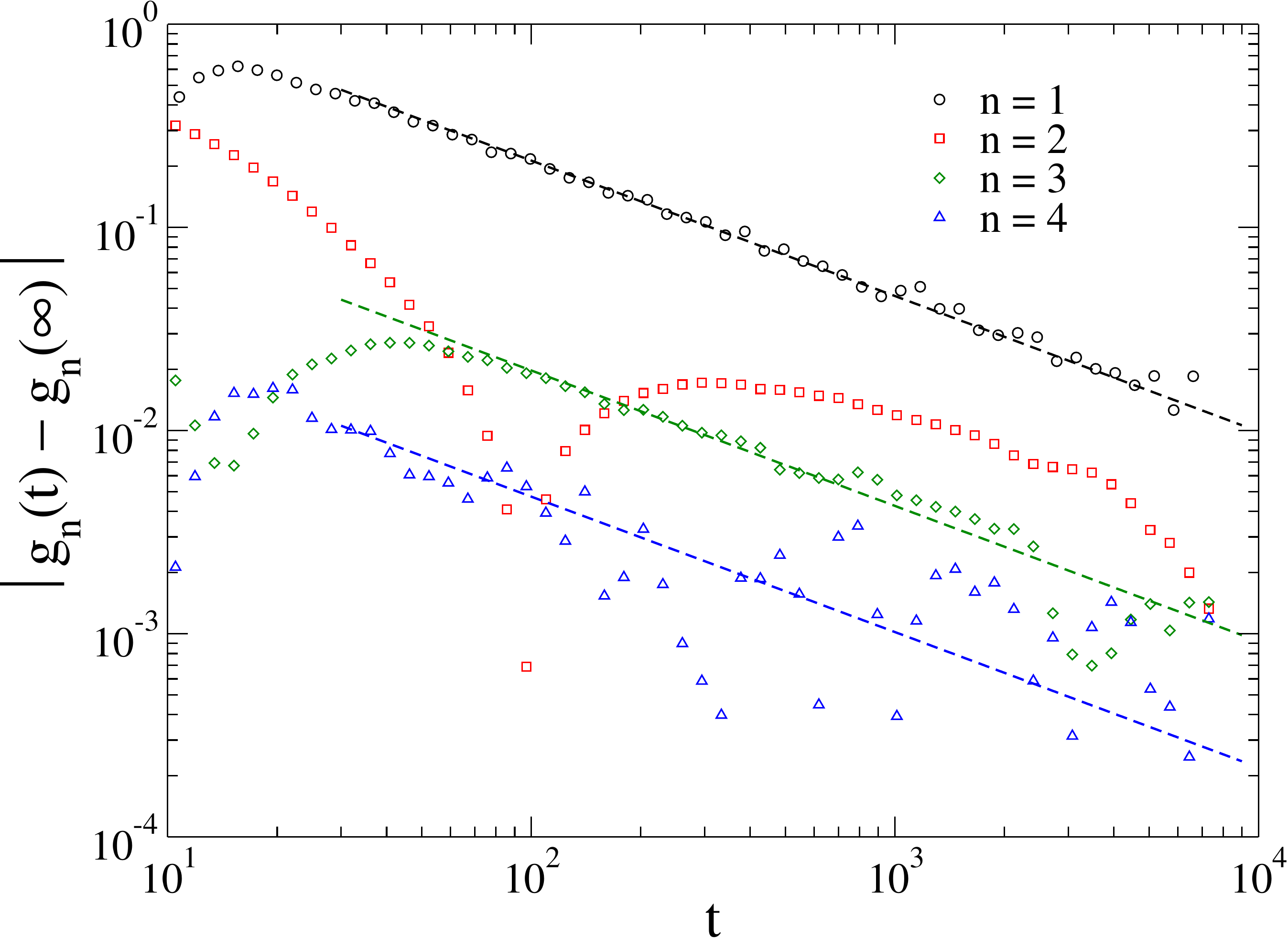}
 \caption{\label{fig:edcorrec} Corrections in scaled cumulants for Eden model.
Dashed lines represent the power law $t^{-2/3}$ as guides to eyes.}
\end{figure}

The main plot of figure~\ref{fig:ed1st} shows the shift
$\lrangle{q}-\lrangle{\chi}$ against time for the Eden model. Using the velocity
$v_\infty=0.51390$ obtained by Takeuchi,  his
result, a decay much faster than $t^{-1/3}$~\cite{TakeuchiJstat}, is reproduced. If the more
accurate estimate $v_\infty=0.51371$ is used instead, a decay close to
$t^{-2/3}$ is observed for short times and is followed by a crossover to a smaller
exponent. The top inset of figure~\ref{fig:ed1st} shows the local exponent
against time. One can clearly see that the scaling exponent has not reached a
stationary value even for our longest simulated times.

Corrections in  $g_n$ are shown in figure~\ref{fig:edcorrec}. The
quantity $g_1$ has an excellent agreement with a $t^{-2/3}$ power law resulting
again in a correction $t^{-1/3}$ in the KPZ ansatz given by equation~(\ref{eq:hpluscorr}).
Thus, a higher order term $t^{-2/3}$ is expected in
$\lrangle{q}-\lrangle{\chi}$ and explains the short-time decay observed in
figure~\ref{fig:ed1st}. Performing a double power law regression
$\lrangle{q}-\lrangle{\chi}=at^{-1/3}+bt^{-2/3}$, an excellent fit is obtained
with positive amplitudes $a=0.50(5)$ and $b=4.0(1)$. This result shows that the
amplitude of the correction $\eta$ is much smaller than the next leading term
resulting a very slow crossover to the asymptotic scaling law $t^{-1/3}$.
Due to the lack of monotonicity, scaling of the corrections in $g_2$ could not be determined
with the present data. The correction in the third cumulant is $t^{-2/3}$. Despite of
large fluctuation, the forth cumulant also decays consistently with 
$t^{-2/3}$.

Even though all parameters in table~\ref{tab:noniso} depend on the non-universal
parameter $\Gamma$, they can be combined in dimensionless 
cumulant ratios that must reflect the universality of $\chi$.
The relative variance $R$, the skewness $S$ 
and the kurtosis $K$ defined by
\begin{equation}
 R = \frac{g_2}{g_1^2} = \frac{\lrangle{\chi^2}_c}{\lrangle{\chi}^2},
\label{eq:R}
\end{equation}
\begin{equation}
 S = \frac{g_3}{g_2^{3/2}} = 
\frac{\lrangle{\chi^3}_c}{\lrangle{\chi^2}_c^{3/2}},
\label{eq:sk}
\end{equation}
and 
\begin{equation}
K = \frac{g_4}{g_2^{2}} = 
\frac{\lrangle{\chi^4}_c}{\lrangle{\chi^2}_c^{2}},
\label{eq:ku}
\end{equation}
must therefore be model-independent.

Table~\ref{tab:cumiso}  shows the cumulant
ratios for all isotropic models investigated. The results are in excellent agreement with
the corresponding GOE ($R_{goe}=1.1046$, $S_{goe}=0.2935$, and $K_{goe}=0.1652$)
for flat models and GUE ($R_{gue}=0.2592$, $S_{gue}=0.2241$, and
$K_{gue}=0.09345$) for the radial Eden model. Scaled height distributions $P(q)$ 
for all isotropic growth models were reported elsewhere~\cite{Oliveira12,TakeuchiJstat}
and, therefore are omitted here. We present scaled HDs for anisotropic growth models
in the following sections.

\begin{table*}[hbt]
\begin{center}
\begin{footnotesize}\begin{tabular}{cccccccc}
\hline\hline
model      &  $g_1$    & $g_2$     & $g_3$       & $g_4$      & $R$      & $S$       & $K$      \\ \hline
RSOS (m=1) & -0.4795(5)& 0.2553(1) & 0.03772(5) & 0.01075(5) & 1.110(2) & 0.292(1) & 0.1649(5) \\ 
RSOS (m=2) & -0.709(1) & 0.5547(4) & 0.1216(2)  & 0.0512(1)  & 1.103(4) & 0.294(2) & 0.1664(5) \\ 
BD         & -1.294(1) & 1.8095(6) & 0.712(1)    & 0.536(3)   & 1.080(2) & 0.292(2)  & 0.163(1)  \\ 
Eden       & -1.770(2) &$\approx$0.81& 0.161(2)  & 0.0607(5)  & 0.26     & 0.22      & 0.093     \\ \hline\hline
\end{tabular}             \end{footnotesize}
\end{center}
\caption{Asymptotic values for $g_n$ for isotropic growth
models. Dimensionless cumulant ratios $R=g_2/g_1^2$, $S=g_3/g_2^{3/2}$ and
$K=g_4/g_2^2$ are also included. Lack of uncertainty in $g_2$ did not allow to
determine errors in dimensionless ratios for Eden model.} 
\label{tab:cumiso}
\end{table*}

\section{Anisotropic radial growth}
\label{sec:anisorad}

Radial interfaces were simulated using Eden models on square lattices.
Two definitions are necessary to describe the models. Periphery sites are
occupied sites with at least one empty nearest neighbor (NN), while growth sites
are empty sites with at least one occupied NN. We investigated two versions. In
the version Eden~A, a growth site is chosen at random and occupied. For each
choice, the time is increased  by $1/N_g$, where $N_g$ is the number of growth
sites. In the version Eden~D on a lattice, a periphery site and one of its NNs
are randomly chosen. If the selected NN is empty, it receives a new particle,
otherwise, the simulation runs to the next step. For each attempt, the time is
incremented by $1/N_p$, where $N_p$ is the number of periphery sites. In all
versions, the growth starts with a single occupied site at the center of the
lattice.

\begin{figure}[ht]
 \centering
 \includegraphics[width=11cm]{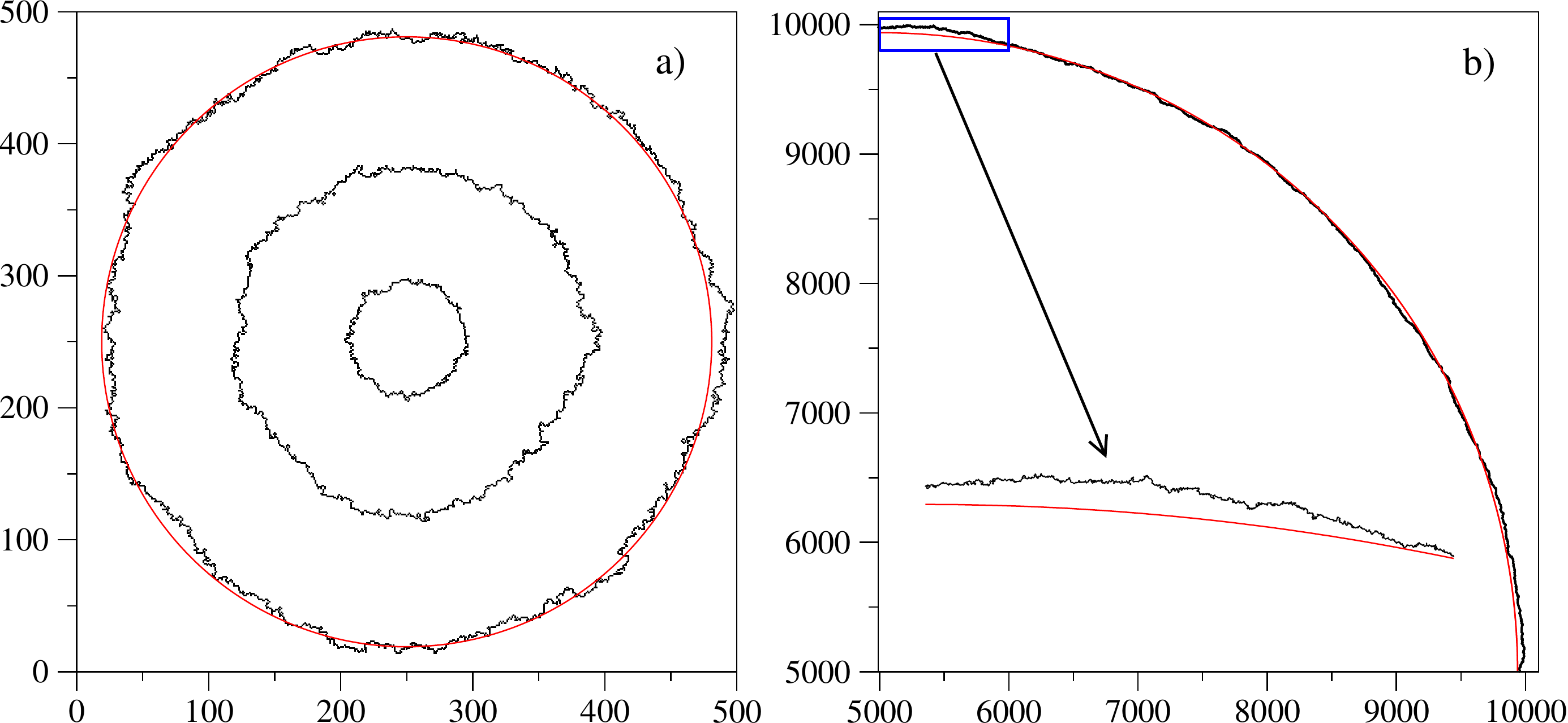}
 \caption{(a) Successive borders of on-lattice Eden~D model for small sizes. 
The circle has the mean radius of the border. (b) First quarter of the border of 
a large on-lattice Eden~D cluster. The inset shows a zoom of the cluster top.}
 \label{fig:padEd}
\end{figure}

The start point to check the equivalence between isotropic and anisotropic
growth is the interface width $w$, defined as the standard deviation of the
radius measured in relation to the center of the lattice. At early times, it
behaves as a power law $w\sim t ^{\beta}$, where the exponent $\beta= 1/3$ is
expected for the KPZ class in $d=1+1$. However, it is well know that Eden
clusters are affected by the lattice-imposed
anisotropy~\cite{Zabolitzky,Paiva07}. More precisely, the axial direction
$\langle 10\rangle$\footnote{Here, we have borrowed the crystallographic
notation where $\langle 10\rangle$ represents the directions of unitary vectors
$\hat{\vec{x}}$,-$\hat{\vec{x}}$, $\hat{\vec{y}}$ and -$\hat{\vec{y}}$.} grows
slightly faster than the diagonal $\langle 11\rangle$ implying that the
dispersion around the mean radius of the border is asymptotically ruled by a
diamond-like shape that produces a crossover from $w\sim t ^{1/3}$ at short
times to $w\sim t$ at long times~\cite{Zabolitzky,Paiva07}.
Figure~\ref{fig:padEd} shows the evolution of an Eden cluster where the faster
growth along $\langle 10\rangle$ can be seen.

Since different directions have different growth velocities, we must focus on
the fluctuations along a fixed direction. The radius on a given direction is
defined as the distance from the origin of the farthest cluster particle along
that direction. Equation~(\ref{eq:hdet}) can be applied for an arbitrary
direction with the parameters $v_\infty$ and $\Gamma$ depending on it. At
a time $t$, we have a collection of radii $\lbrace r_1(t),r_2(t),\cdots
r_N(t)\rbrace$ along a given direction that are obtained from an ensemble of $N$
independent simulations. 
We analyzed axial and diagonal 
directions that correspond to fastest and slowest growth directions in square
lattices, respectively. We simulated $2.5\times 10^6$ clusters of diameter 
$5\times 10^3$. Therefore, our statistics is performed with $10^7$ 
points for each analyzed direction. A typical simulation of Eden~A and D
takes about 3~s and 7~s, respectively, in a CPU Intel Xeon 3.20GHz.  

\begin{figure}[ht]
 \centering
 \includegraphics[width=9cm]{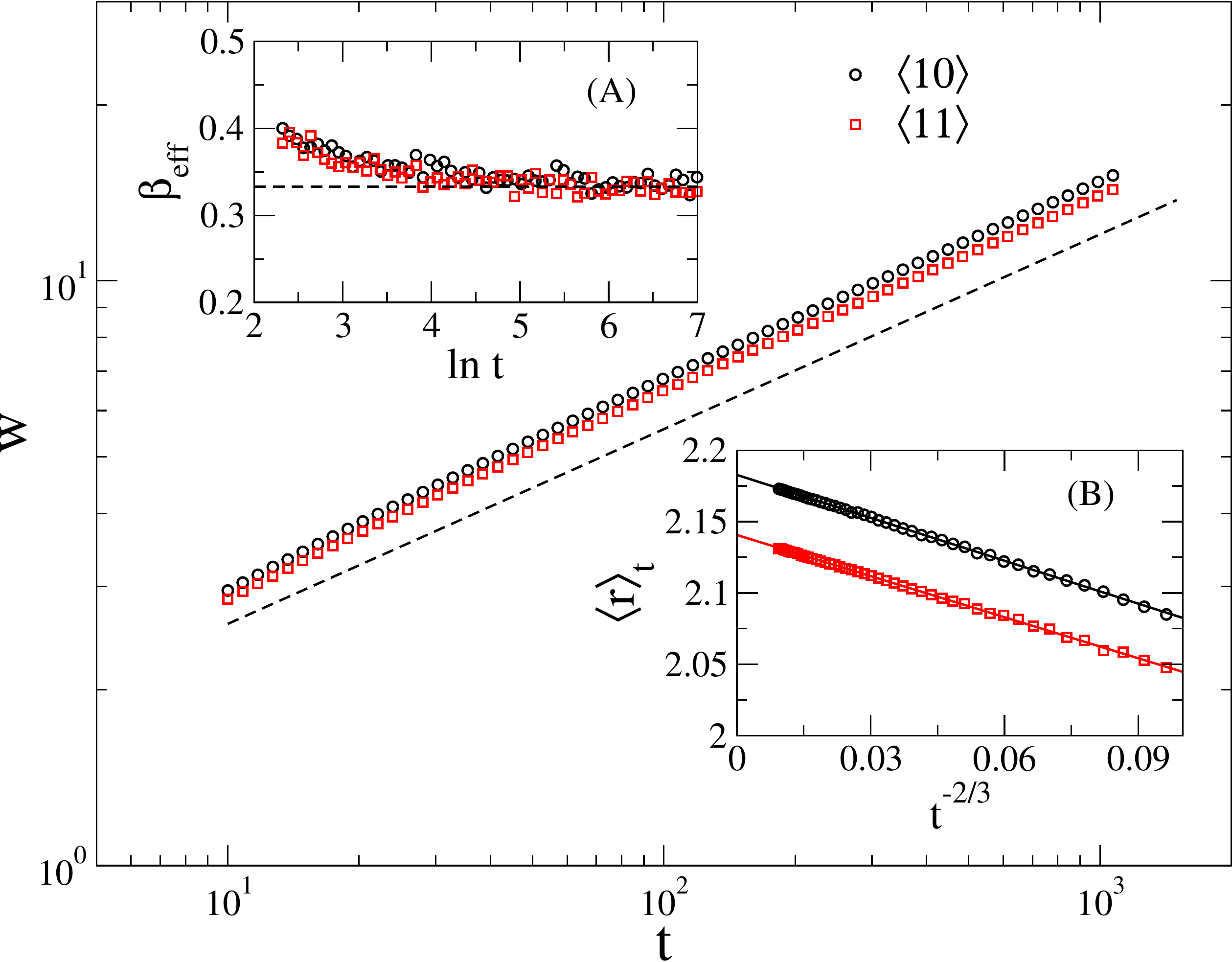}\\~\\
 \caption{Interface width against time for on-lattice Eden~A model. 
Inset (A): Effective growth exponent obtained  from the  
derivative of $\ln w$ versus $\ln t$. Inset (B): Interface velocity 
against $t^{-2/3}$.}
 \label{fig:rugEdA}
\end{figure}

Figure~\ref{fig:rugEdA} shows the interface width against time for Eden~A
computed for distinct directions. Interface width along direction $\lrangle{10}$
is larger than along direction $\lrangle{11}$ ($\Delta w \approx$ 6\%), but they
follow  scaling laws in time with growth exponents $\beta_{10} = 0.335(7) $ and
$\beta_{11} = 0.332(7)$, respectively, in excellent agreement with the KPZ
universality class.  Growth exponents for Eden~A and D are shown in
table~\ref{tab:uni}. The inset B of figure~\ref{fig:rugEdA} shows the analysis
to determine the interface velocity for Eden~A simulations. A higher velocity 
along axial direction is evident.

\begin{figure}[ht]
 \centering
 \includegraphics[width=9cm]{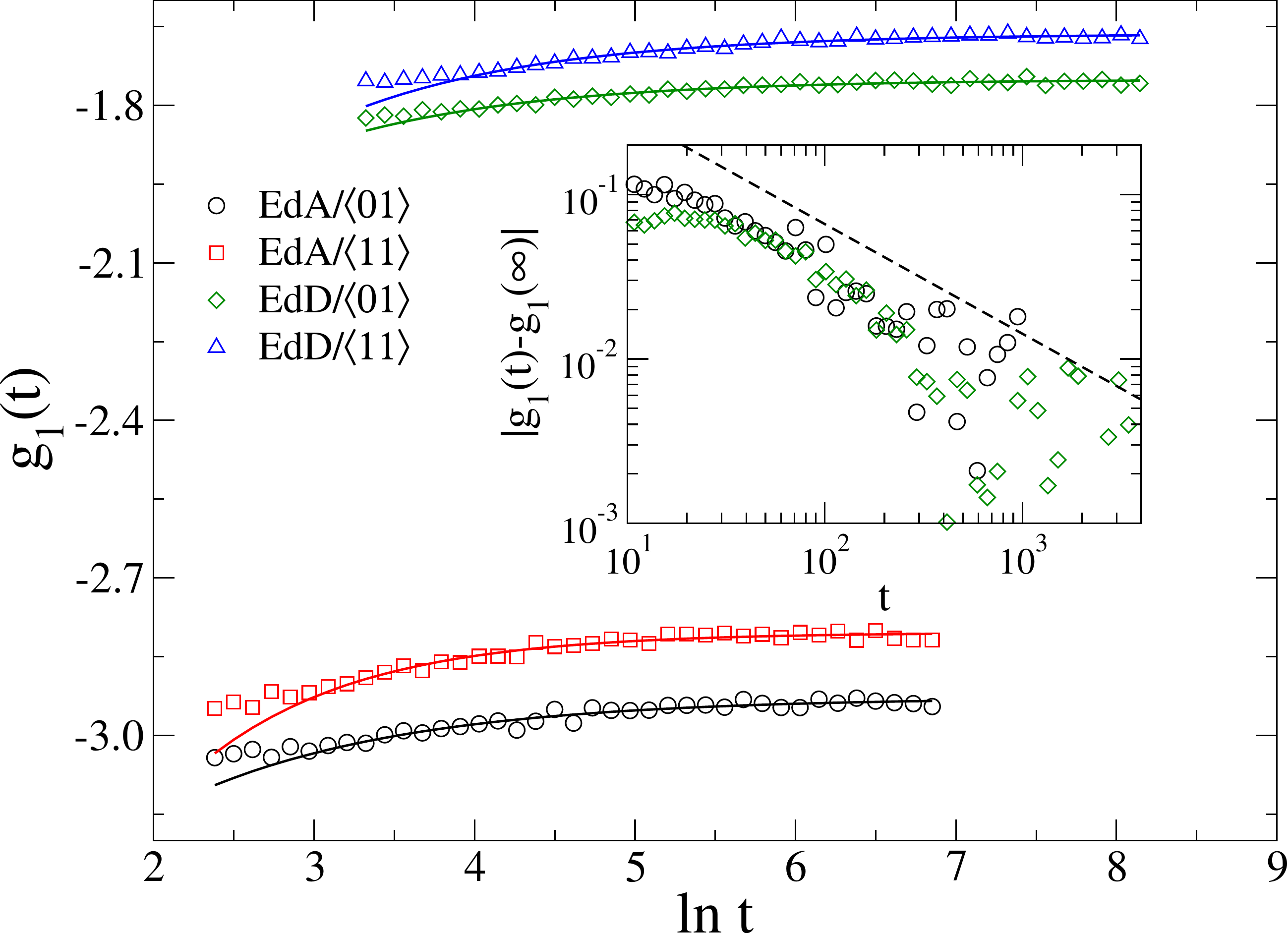}
 \caption{Determination of the non-universal 
quantity $\Gamma^{1/3}\lrangle{\chi}$ for on-lattice Eden models. 
Solid lines are non-linear regressions
used to extrapolate $g_1$. The inset
shows the correction in $g_1$ against time. Dashed line has slope -2/3.}
\label{fig:g1_eden}
\end{figure}

\begin{figure}[t]
 \centering
 \includegraphics[width=9cm]{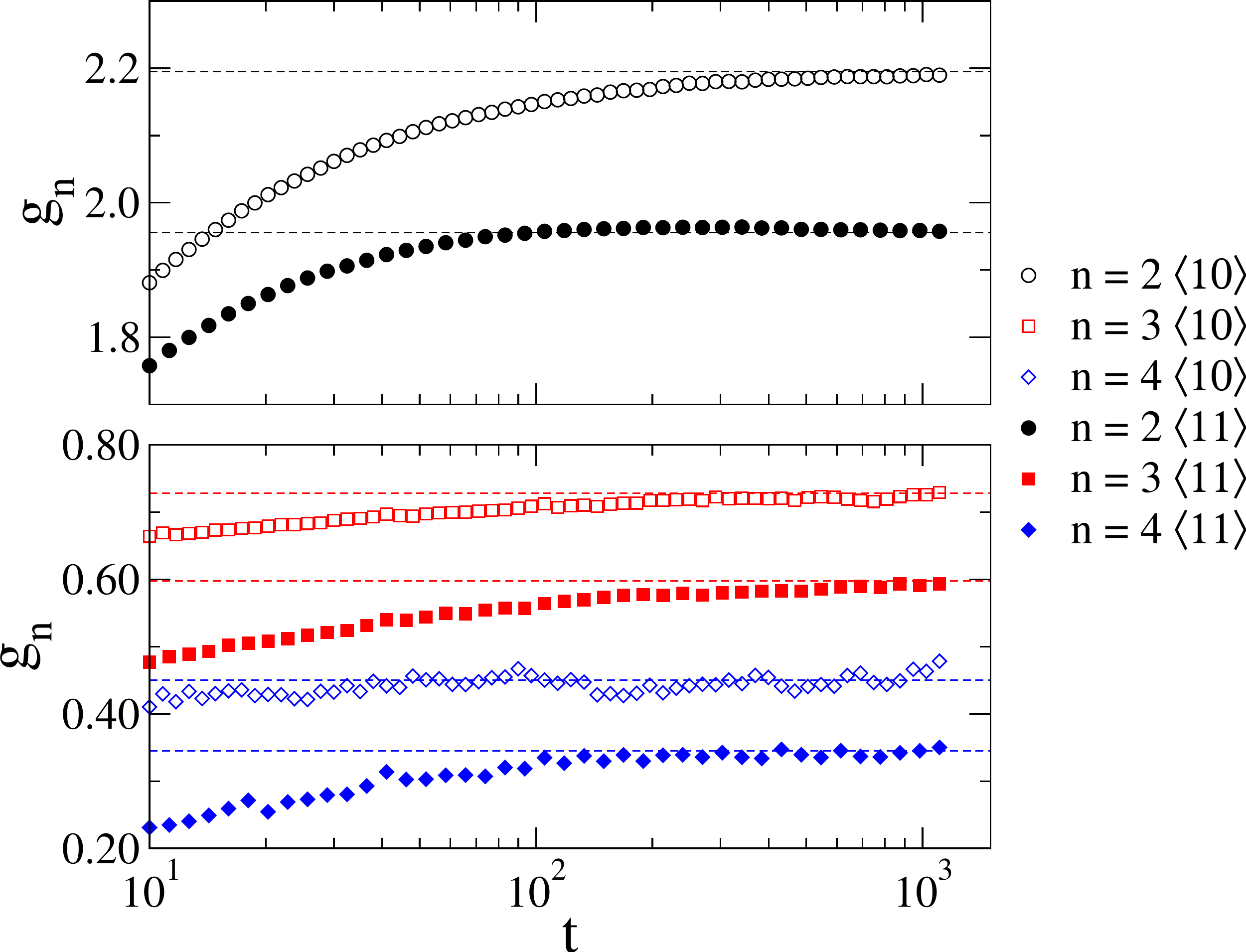}
\caption{Scaled cumulants $g_n = \lrangle{r^n}_c / (s_\lambda^n t^{n/3})$ for
on-lattice Eden~A along directions $\lrangle{10}$ (open symbols) and
$\lrangle{11}$ (filled symbols). Lines are extrapolations to
$t\rightarrow\infty$.}
 \label{fig:cumEdAvst}
\end{figure}

\begin{figure}[t]
 \centering
 \includegraphics[width=8cm]{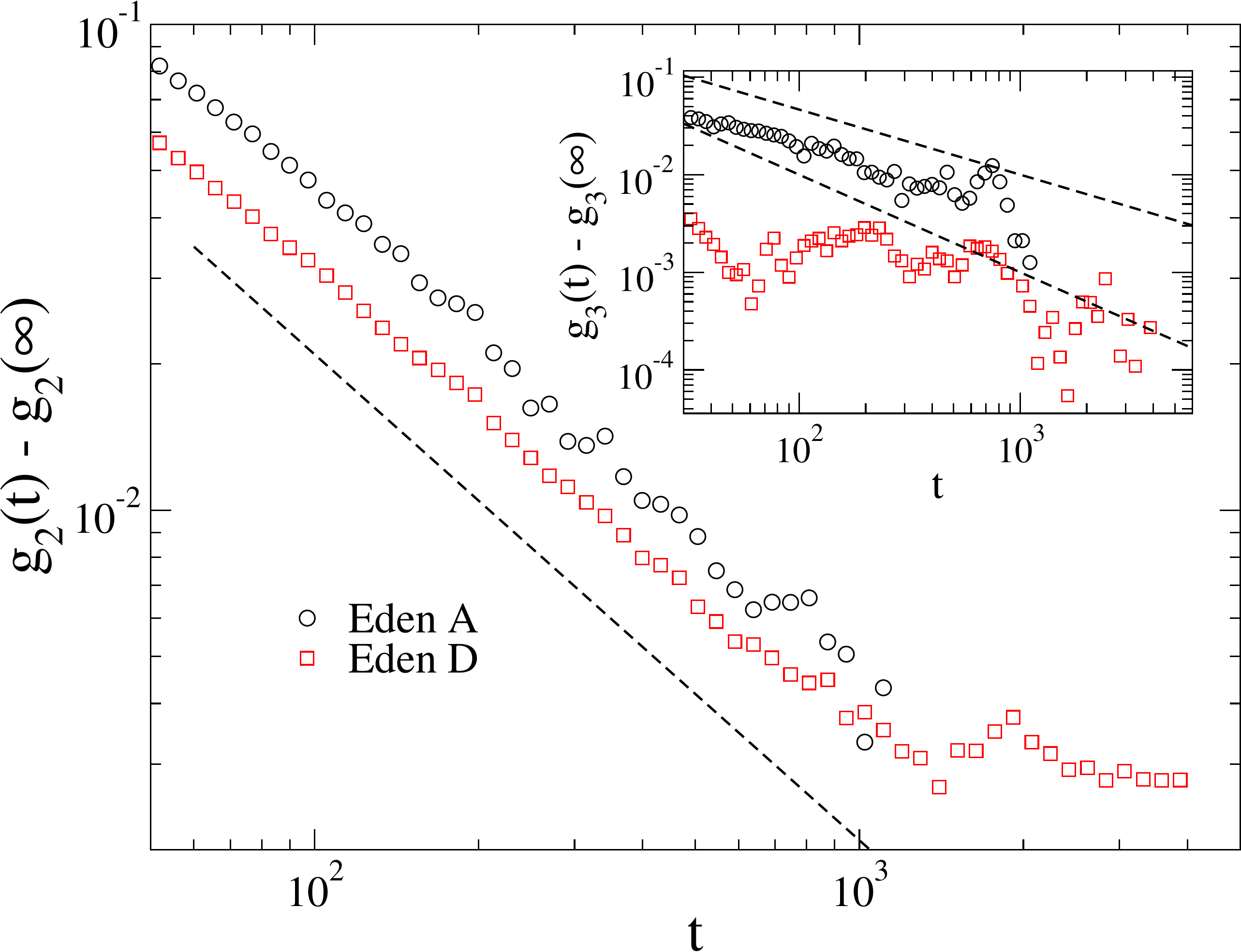}
 \caption{Corrections in the second cumulant for on-lattice Eden models along direction
$\lrangle{10}$. The dashed line is a power law with exponent $-1$. Inset:
Corrections in the third cumulant. The dashed lines are power laws with exponents
$-2/3$ and $-1$.}
 \label{fig:corrEd}
\end{figure}

Corrections in $g_1$ are shown in figure~\ref{fig:g1_eden} and are consistent
with a decay $t^{-2/3}$ implying that the next leading term in
equation~(\ref{eq:hpluscorr}) is $t^{-1/3}$, in analogy with isotropic growth
models described in sections~\ref{sec:isoflat} and~\ref{sec:isorad}. The
scaled cumulants against time for Eden~A are shown in
figure~\ref{fig:cumEdAvst}.  Along the  direction $\lrangle{10}$, we have found
corrections consistent with $t^{-1}$ for the second cumulant in both Eden A and
D. Along direction $\lrangle{11}$, $g_2$ has a slight non-monotonicity that
complicates the extrapolation to long times. It was not possible to accurately
resolve the scaling of the corrections for higher order cumulants but we have
found that they do not decay slower than $t^{-2/3}$, as illustrated in
figure~\ref{fig:corrEd}. This indicates that the correction $\eta$ is
statistically independent of $\chi$ since, otherwise, corrections  would decay
as $t^{-1/3}$ for all higher order cumulants~\cite{TakeuchiJSP12,Ferrari}.
Notice that many features of our current on-lattice simulations were observed
for the isotropic growth, where $t^{-2/3}$ was found for $g_1$, $g_3$ and $g_4$
and a non-monotonicity for $g_2$. Table~\ref{tab:nonaniso} shows the non-universal
parameters $v_\infty$ and $g_n$ obtained in our simulations.  

One can see in table~\ref{tab:nonaniso} that the estimates for $\Gamma_1$  are
slightly larger than those for $\Gamma_2$ for all investigated models, as
observed for ballistic deposition on flat substrates. In the rest of this work,
we use $g_1$ to investigate the scaling and universality of the models since it
yields reliable estimates of $\Gamma$ as discussed in section~\ref{sec:isoflat}.

\begin{table}[ht]
\caption{Non-universal parameters of several anisotropic growth models with curved
geometry. Acronyms: EdA (Eden~A); EdD (Eden~D); WRSOS (wedge restricted
solid-on-solid); BDD (ballistic deposition droplet). 
The missing uncertainties for EdA/$\lrangle{11}$ and EdD/$\lrangle{11}$ are due to the lack of
monotonicity in the second cumulant of the respective data.}
\label{tab:nonaniso}
\begin{footnotesize}
\begin{tabular}{ccccccccc}
\hline
Model               & $v_\infty$         & $g_1$     & $g_2$       &$g_3$     &$g_4$     & $\Gamma_1$  & $\Gamma_2$ & $\lrangle{\eta}$  \\ \hline
EdA/$\lrangle{10}$  &  2.1824(3)  & -2.927(3) &  2.195(1)   &0.728(2)  &0.45(1)   & 4.51(1)     & 4.43(1)    & 1.7(1)\\
EdA/$\lrangle{11}$  &  2.1401(3)  & -2.807(3) &  1.956      &0.598(2)  &0.345(9)  & 3.98(1)     & 3.73       & 1.1(1)\\ 
EdD/$\lrangle{10}$  &  0.6186(1)  & -1.754(2) &  0.779(1)   &0.1535(8) &0.057(1)  & 0.971(3)    & 0.938(2)   & 1.6(1)\\
EdD/$\lrangle{11}$  &  0.61055(5) & -1.677(2) &  0.695      &0.127(2)  &0.044(1)  & 0.85(1)     & 0.79       & 1.1(1)\\
WRSOS               &  0.41903(1) &  -1.115(4) &  0.3233(2)  &0.0418(2)&0.0095(4) & 0.250(3)    & 0.251(1)   & -0.7(1)\\
BDD                 &  2.13984(2) & -3.023(9) &  2.331(5)   &0.802(6)  &0.49(1)   & 4.97(4)     & 4.85(2)    &1.3(1)\\ \hline 
\end{tabular}
\end{footnotesize}
\end{table}

\begin{table}[ht]
\caption{Universal quantities for several anisotropic models with curved geometry. Acronyms
as in table~\ref{tab:nonaniso}. Cumulant ratios are defined by equations 
(\ref{eq:R})-(\ref{eq:ku}).
The KPZ cumulant ratios correspond to the GUE distribution were taken from reference~\cite{PraSpo1}.
The lack of uncertainties in the second cumulant of EdA/$\lrangle{11}$ and EdD/$\lrangle{11}$ models
did not allow the error propagation in the cumulant ratios.}
\label{tab:uni}
\begin{center}
\begin{tabular}{ccccc}
\hline
Model              & $\beta$        & $R$      & $S$      & $K$\\ \hline
EdA/$\lrangle{10}$ & 0.335(7)       & 0.256(1) & 0.224(1) & 0.093(2)\\
EdA/$\lrangle{11}$ & 0.332(7)       & 0.25     & 0.22     & 0.090\\ 
EdD/$\lrangle{10}$ & 0.338(9)       & 0.253(1) & 0.223(2) & 0.094(2) \\
EdD/$\lrangle{11}$ & 0.334(10)      & 0.25     & 0.22     & 0.091 \\
WRSOS              & 0.3326(5)      & 0.260(2) & 0.227(1) & 0.091(4)  \\ 
BDD                & 0.335(2)       & 0.255(2) & 0.225(2) & 0.090(2)  \\ 
KPZ                & 1/3            & 0.2592   & 0.2241   & 0.09345 \\ \hline
\end{tabular}
\end{center}
\end{table}

\begin{figure}[ht]
 \centering
 \includegraphics[width=9cm]{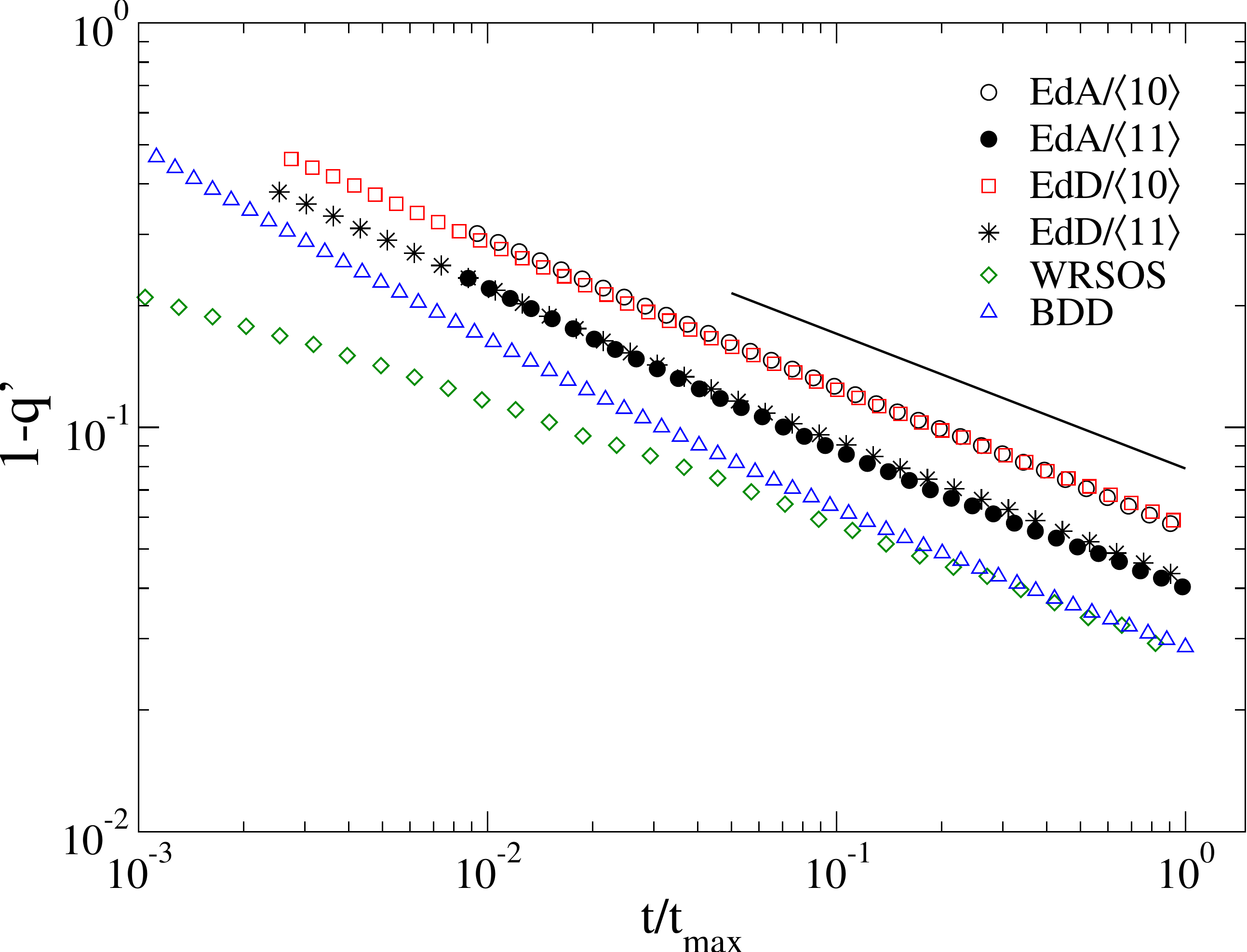}
 \caption{Scaling law describing the shift in the mean of scaled surface
distributions for several anisotropic KPZ models. The time was scaled by the maximum
growth time reached in each model in order to improve visibility.
The solid line represents the power law $t^{-1/3}$. 
Acronyms as in table~\ref{tab:nonaniso}.}
 \label{fig:shift1}
\end{figure}

Due to the anisotropy, the surfaces studied
here do not have translational invariance so that two-point local quantities
are not well-defined. So, the parameter $A$ and, consequently, 
$\Gamma$ cannot be directly measured if cumulants of $\chi$ are unknown. We then define 
\begin{equation}
q' = \frac{r-v_\infty t}{s_\lambda g_1 t^{1/3}} 
\equiv \frac{q}{\lrangle{\chi}}
\label{eq:qprime}
\end{equation}
such that $\lrangle{q'}\rightarrow 1$ for $t\rightarrow\infty$. We can obtain
the scaling law ruling the shift in the mean by plotting $1-\lrangle{q'}$ 
against time using only the directly measurable parameters $v_\infty$ and $g_1$.
The results are shown in figure~\ref{fig:shift1}, where we have obtained
corrections very consistent with $t^{-1/3}$, already observed in many other
systems~\cite{TakeSano,TakeuchiSP,SasaSpo1,Alves11,Ferrari}. The slow crossover
to $t^{-1/3}$ observed for  off-lattice simulations of
Eden~D~\cite{TakeuchiJstat} was not observed in the on-lattice version.
Moreover, our previous off-lattice simulations of Eden~B\footnote{The difference
between Eden~D and Eden~B is that any nearest neighbor can be chosen in the
former while only the empty ones are eligible in the latter.} also yielded a
shift consistent with $t^{-1/3}$~\cite{Alves11}. An important remark is that the
directly measurable amplitude in the scaling $1-\lrangle{q'}\simeq -
\lrangle{\eta}/g_1 t^{-1/3}$ also yields a way to obtain an estimate of
$\lrangle{\eta}$.  The values of $\lrangle{\eta}$ for anisotropic growth 
models are shown in table~\ref{tab:nonaniso}.

The dimensionless cumulant ratios obtained for both Eden
models are shown in table~\ref{tab:uni}.  All ratios are universal and show
remarkable agreement with GUE~\cite{PraSpo1}. Notice that using dimensionless
cumulant ratios as equations~(\ref{eq:R}), (\ref{eq:sk}) and (\ref{eq:ku}), all
cumulants of $\chi$ can be determined as functions of a single cumulant. This
reasoning also applies to systems where $\chi$ is unknown raising a general
strategy to probe universality in the interface distributions.

\begin{figure}[ht]
 \centering
 \includegraphics[width=8cm]{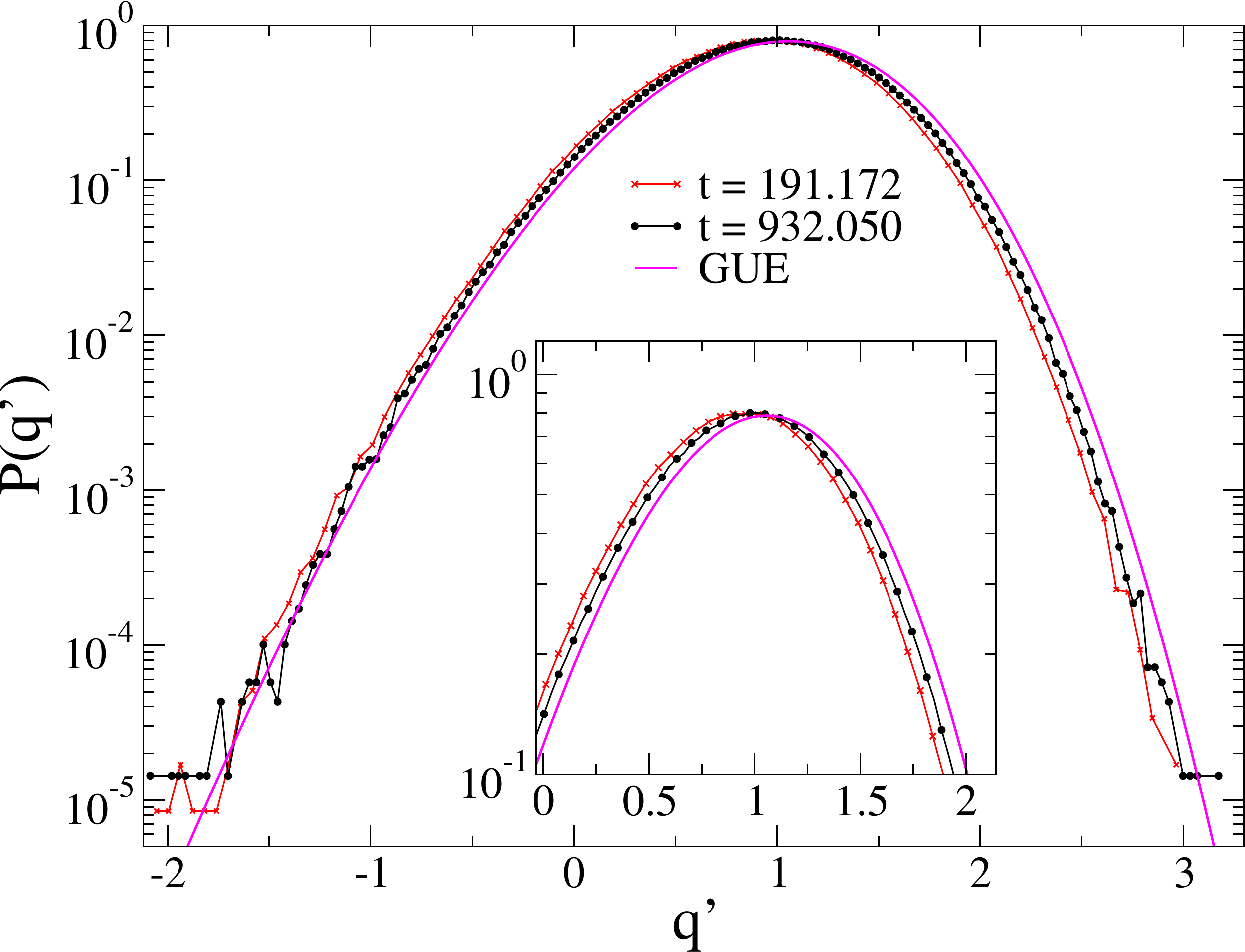} 

\includegraphics[width=8cm]{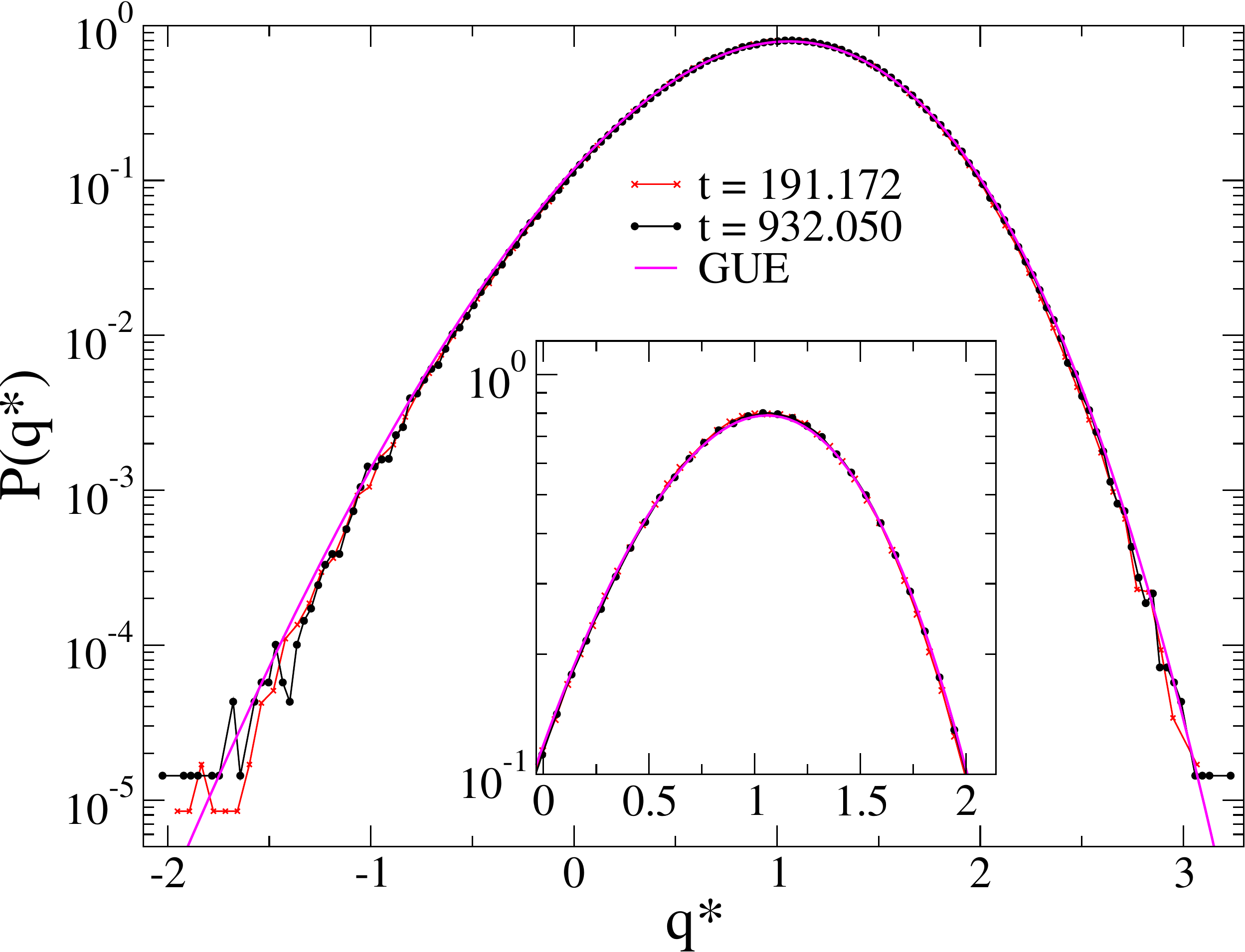}
 \caption{Top: Distributions for Eden~A model at different growth times  scaled 
accordingly equation~(\ref{eq:qprime}).  
Bottom: Same distributions rescaled using equation~(\ref{eq:qstar}) that 
includes the shift $\lrangle{\eta}=1.7$. Insets show zooms around the 
peaks.}
 \label{fig:pofqEdA}
\end{figure}

The top panel of figure~\ref{fig:pofqEdA} compares the distributions scaled
accordingly equation~(\ref{eq:qprime}) for Eden~A in the
direction~$\lrangle{10}$ at different  times with the GUE distribution scaled to
a mean 1, $\chi^*=\chi/\lrangle{\chi}$. The distributions converge to GUE as
time evolves but the shift is evident. If the estimated shift  $\lrangle{\eta}
=1.7(1)$ is explicitly included by
defining
\begin{equation}
 q^* = \frac{r-v_\infty t-\lrangle{\eta}}{s_\lambda g_1 t^{1/3}},
\label{eq:qstar}
\end{equation}
an excellent collapse between $P(q^*)$ and $P(\chi^*)$ is found
(figure~\ref{fig:pofqEdA}, bottom). Similar results are found for all
investigated models. 

\section{Droplet growth}
\label{sec:droplet}

A paradigmatic model for the growth of a droplet morphology is the 
PNG model, where islands are randomly nucleated over
existing ones and, once nucleated, they starts to grow laterally with constant
velocity. If we consider an initial island that grows indefinitely from a single
nucleation at the origin, the PNG model has an exact solution with an asymptotic
droplet shape, where the height at the origin is given by equation~(\ref{eq:hdet}) with
exactly known parameters~\cite{PraSpo1,PraSpo2}. PNG model simulations with
droplet geometry  have been performed to be compared with analytic
results~\cite{PraSpo1,SchehrEPL}. Other models and
initial conditions leading to droplet interfaces has also been investigated~\cite{SasaSpo1,johansson,Prolhac}.

\begin{figure}[ht]
 \centering
 \includegraphics[width=7cm]{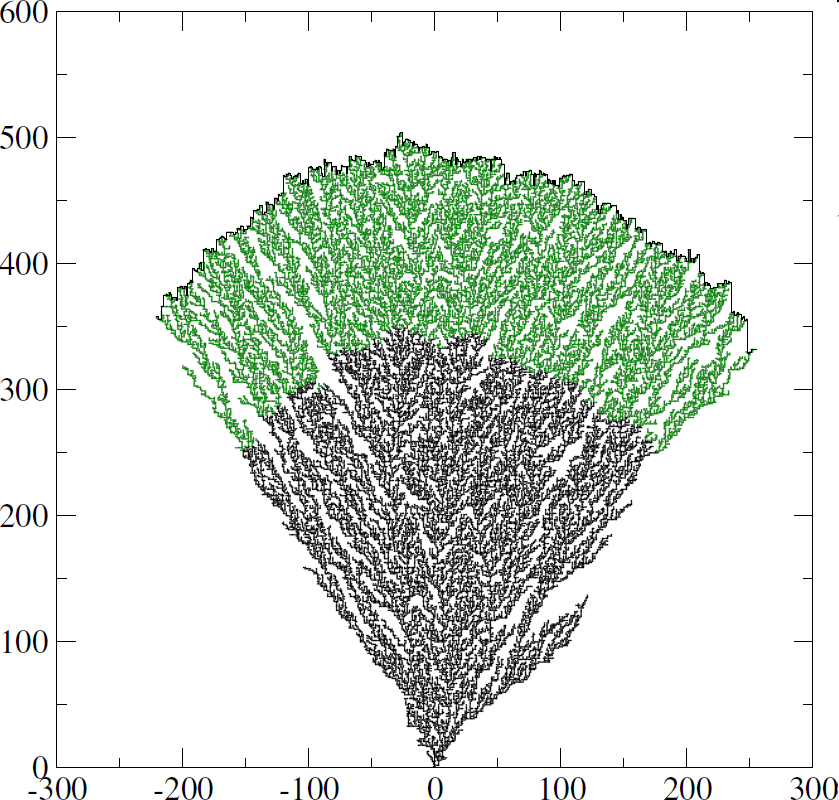}
 \caption{Ballistic deposition droplet after deposition of 56000 particles. The
first half of deposited particles are depicted in black and the rest in green.
The solid line represents the interface.} 
\label{fig:bdd}
\end{figure}

Droplet geometry can also be investigated in more complex models for which
analytical results are currently unavailable as, for example, the BD model \cite{Robledo11}. 
The growth of ballistic deposition droplets (BDD) starts with a single particle stuck to
the origin of the system. Particles move ballistically along direction $-y$. If
a particle visits a NN site of the aggregate, it
irreversibly attaches to this position and becomes part of the aggregate. Let
$L(t)$ be the size of the active region defined as the set of $x$-coordinates
where a growth can be tried. When a particle deposition is tried the time is
increased by $\Delta t = 1/L(t)$. 
Figure~\ref{fig:bdd} shows a typical ballistic deposition droplet. 

The non-universal parameters depend on the position in the active region in
analogy to the direction dependence in the Eden growth, whereas cumulant ratios
and growth exponent do not. An ensemble of $2\times 10^7$ samples were used to
compute statistics. Growth of  a cluster for $t=10^4$ ($\sim10^8$ particles)
takes about $2$~s in an CPU Intel Xeon  3.20GHz.  For sake of conciseness, we
present only results for the height fluctuations at the origin. 

\begin{figure}[ht]
 \centering
 \includegraphics[width=9cm]{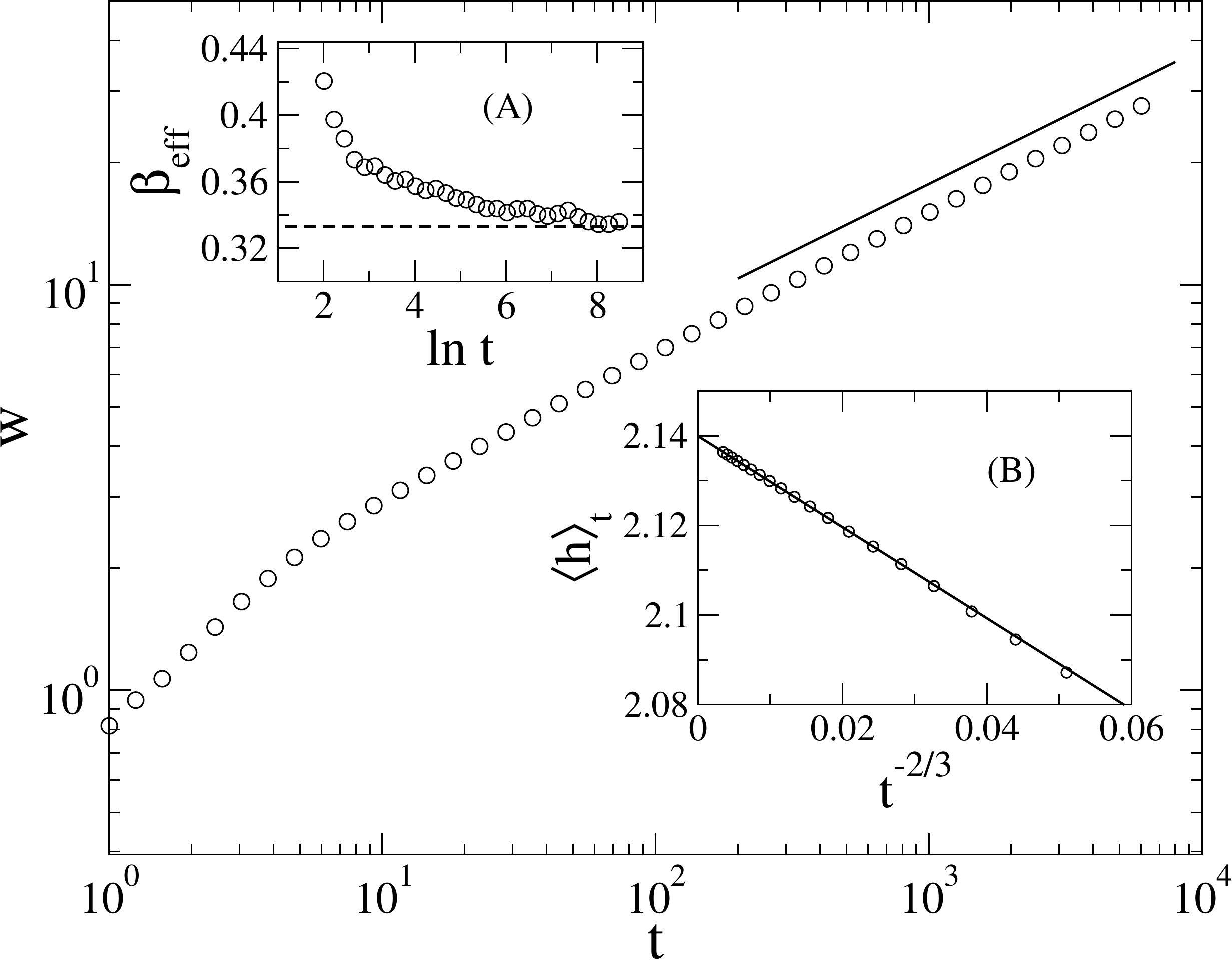}
 \caption{Interface width against time determined at center of the
BDD model. Solid line is a power law $t^{1/3}$. The inset (A)
shows the effective growth exponent while inset (B)
shows the interface velocity against $t^{-2/3}$.}
 \label{fig:rugBDD}
\end{figure}

\begin{figure}[ht]
 \centering
 \includegraphics[width=9cm]{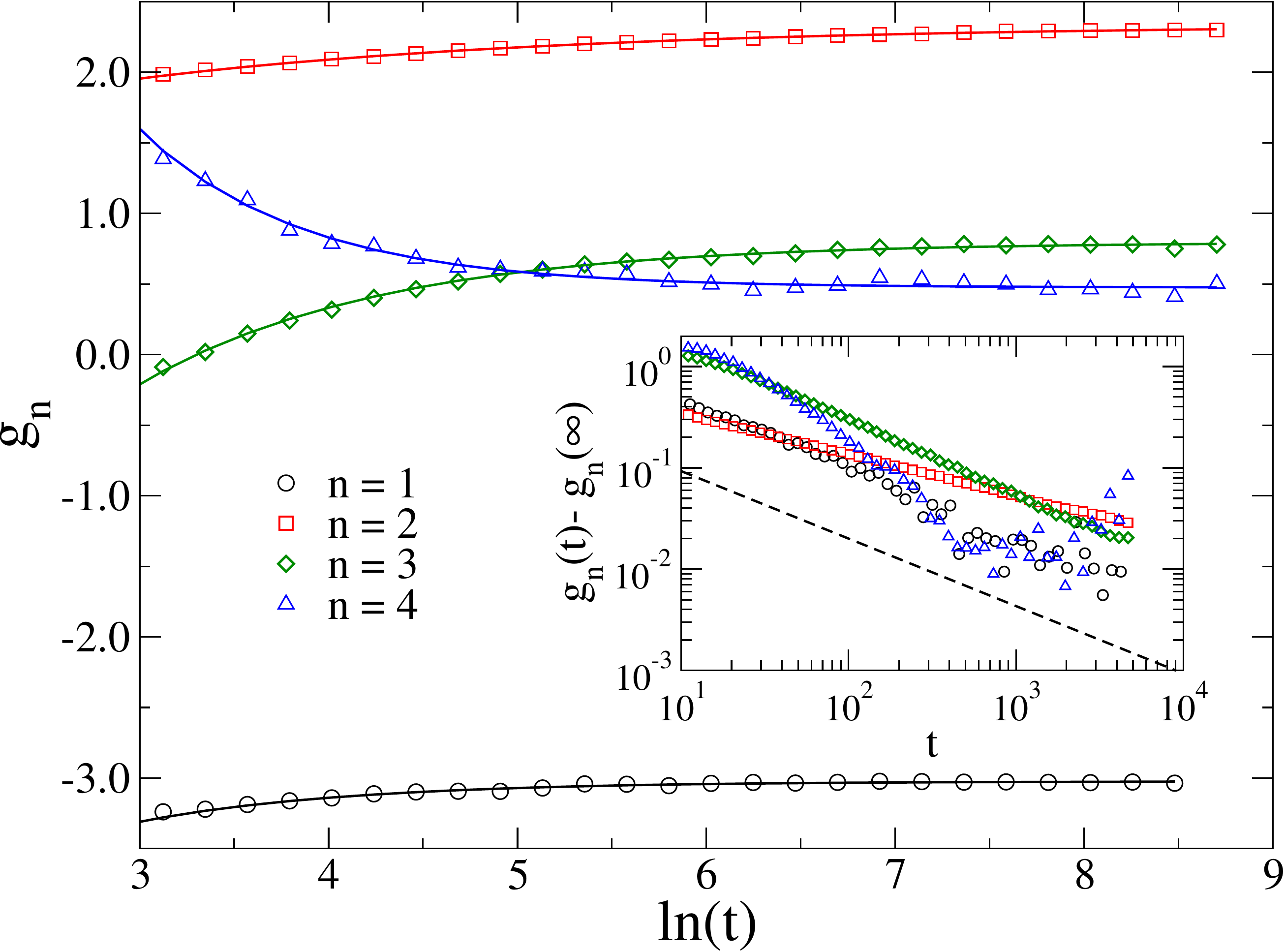}
 \caption{Semi-logarithmic plot of
$g_n$ against time for BDD model. Solid lines are 
the non-linear regression used to extrapolate the asymptotic values. 
Inset: Difference $g_n(t)-g_n(\infty)$ 
against time. The dashed line represents a decay $t^{-2/3}$.}
 \label{fig:gnBDD}
\end{figure}

The curves used to determine the non-universal parameters (table~\ref{tab:nonaniso})
are shown in figures~\ref{fig:rugBDD} and~\ref{fig:gnBDD}. The interface
velocity has the usual behavior illustrated in the inset (B) of
figure~\ref{fig:rugBDD}. However, BDD growth has strong corrections in the 
growth exponent as confirmed in the inset (A) of figure~\ref{fig:rugBDD}, in
analogy to its flat counterpart~\cite{Oliveira12,Bahman11}. The correction in
the mean converges in the usual way ($t^{-1/3}$) as shown in
figure~\ref{fig:shift1}.  Notice the presence of a crossover from a faster
initial decay to the regime $t^{-1/3}$, analogous to the crossover observed for
the isotropic Eden~D model. Similarly to the isotropic growth, corrections in $g_2$ 
decays with an unusually small exponent $t^{-0.45}$. 

The quantity  $g_1$ approaches its asymptotic value as $t^{-2/3}$,  the same
correction obtained for flat simulations of isotropic growth
models and for all versions of the anisotropic Eden model
presented in section~\ref{sec:anisorad}. The cumulant of order $n=3$ also has a
correction very close to $t^{-2/3}$ while the cumulant of order $n=4$ has a
correction faster than $t^{-2/3}$. In particular, the correction observed in
$g_3$ is slower than that observed for the flat case,
demonstrating the intricate and non-universal scenarios involving corrections in
cumulants of order $n\ge 2$. The inset of figure~\ref{fig:gnBDD} shows the 
corresponding corrections in $g_n$.  
Finally, the growth exponent and dimensionless cumulant ratios converge to the values
expected for the KPZ class as shown in table~\ref{tab:uni}.

\begin{figure}[ht]
 \centering
 \includegraphics[width=9cm]{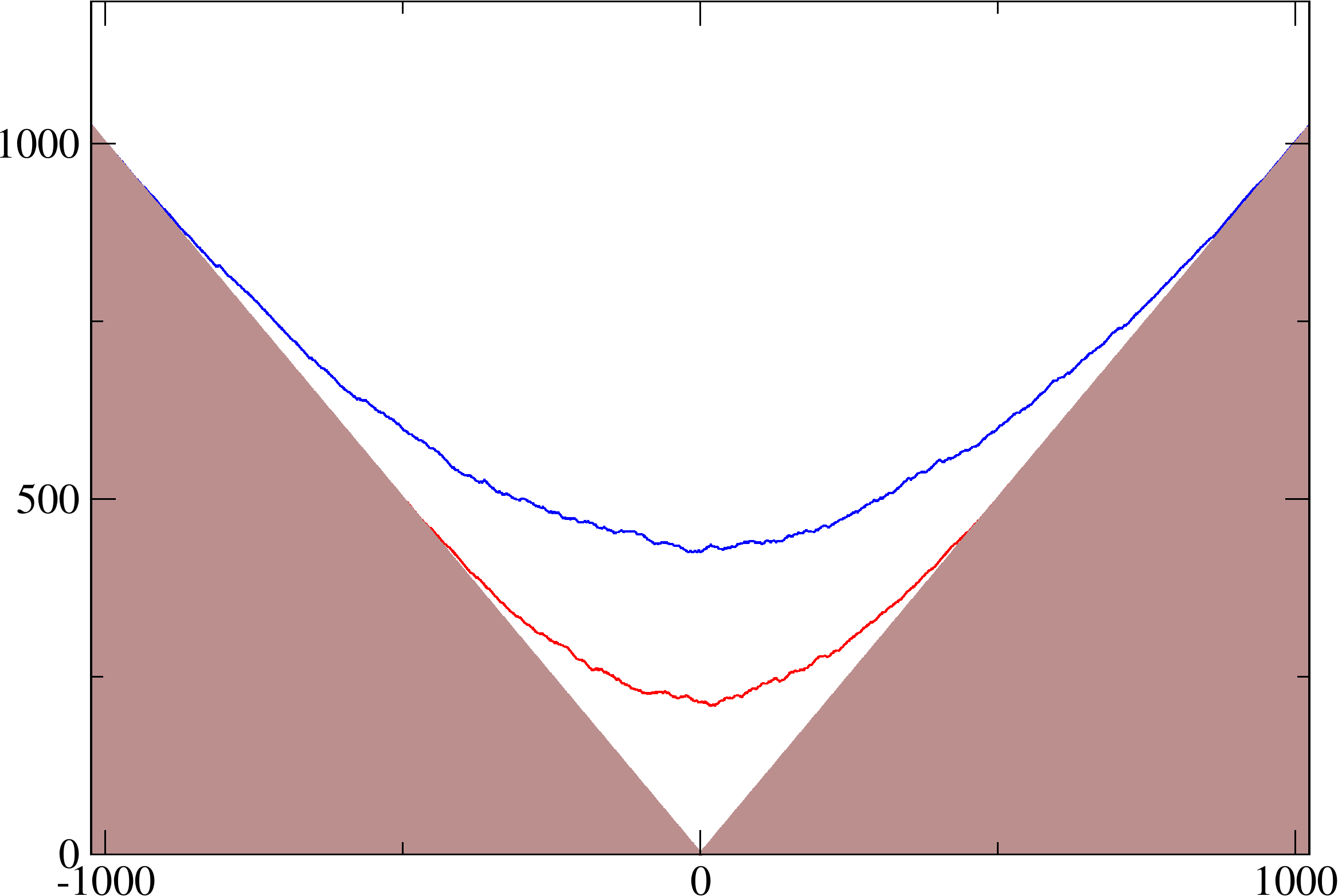}
 \caption{Typical interfaces obtained for WRSOS model 
 for deposition times $t=512$ (lower)  and $t=1024$ (upper) .}
 \label{fig:kkpad}
\end{figure}

We also considered the restricted solid-on-solid (RSOS) growth model~\cite{kk}
with a wedge initial condition, the WRSOS model. At each time step, a site in
the growing zone (of size $L(t)$) is randomly selected and its height increased
by 1 if the constraint $\Delta h = |h(j,t)-h(j\pm 1,t)|\le 1$ is satisfied,
otherwise, the deposition attempt is refused. The time is incremented by $\Delta
t = 1/L(t)$ for each attempt. A wedge initial condition $h(j,0)=|j|$, obeying
$\Delta h = 1$, was considered. This initial condition implies a droplet growth
since the radius of the growing zone increases with average velocity  $1$  while
in the center ($\lrangle{10}$ direction) it has a smaller velocity (Tab.
\ref{tab:nonaniso}). Typical interfaces for distinct times are shown in
figure~\ref{fig:kkpad}.

\begin{figure}[ht]
 \centering
 \includegraphics[width=9cm]{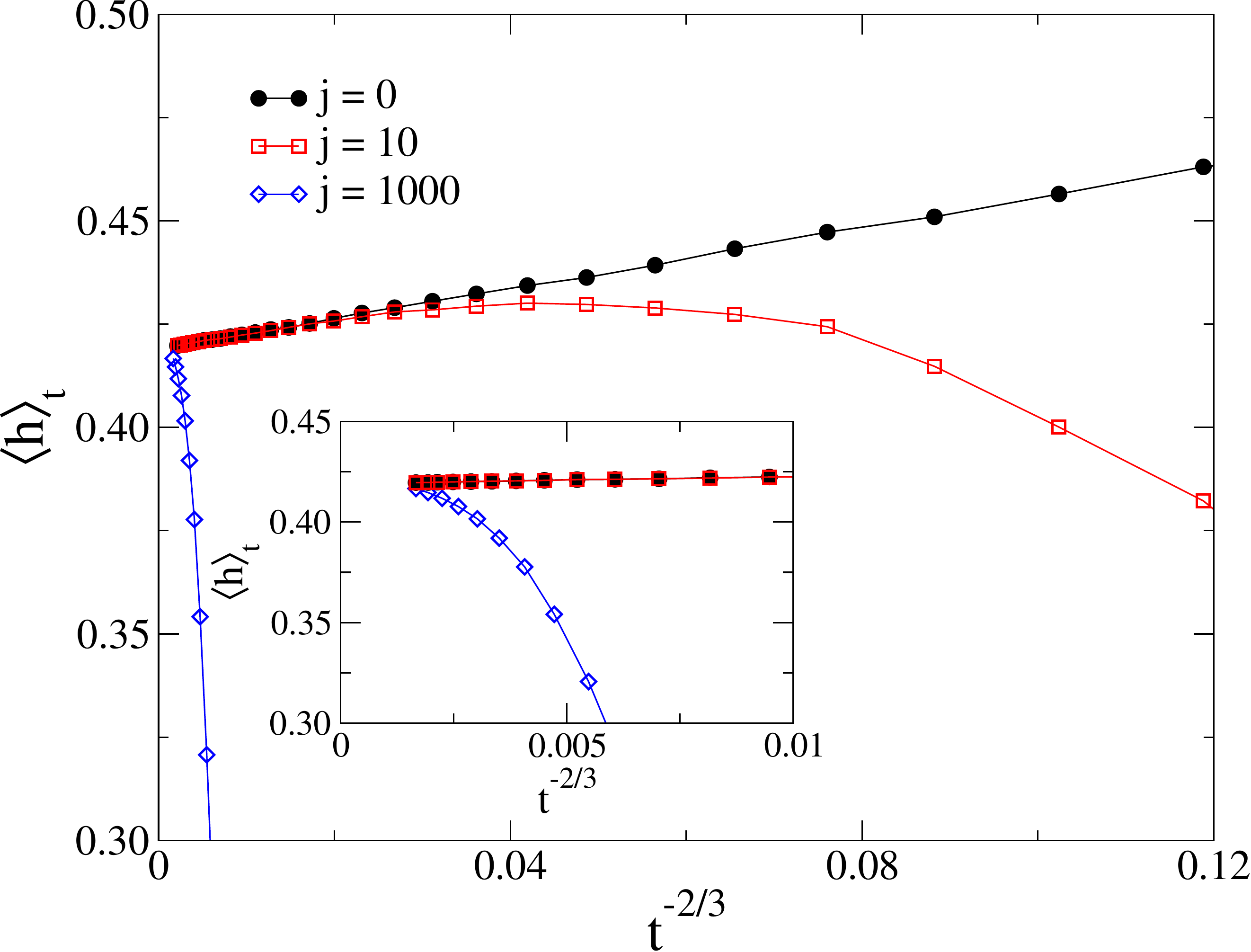} \\
 \caption{Interface velocity at different positions of WRSOS growth. The inset
shows a zoom to show the asymptotic velocity.}
 \label{fig:kkv}
\end{figure}

Figure~\ref{fig:kkv} shows the interface velocity at $j=0$, $10$, and $1000$.
They converge to the same asymptotic value observed for the flat
geometry~\cite{Oliveira12}, but differ at short times. This means a macroscopic
shape, around the droplet center, that asymptotically moves with constant velocity. The macroscopic shape is
almost perfectly fitted by a parabola $h(x)=h(0)+ax^2$ in analogy with the
solution of the KPZ equation with a sharp wedge initial
condition~\cite{SasaSpo1}. The surface statistics was computed using an ensemble
of $5\times10^7$ samples. A typical run up to $t=3000$ takes about 0.2~s in a
CPU Intel Xeon  3.20GHz.  Non-universal parameters obtained for the height
fluctuations at the origin $j=0$ are shown in table~\ref{tab:nonaniso}. The universal
quantities shown in table~\ref{tab:uni} exhibit an excellent agreement with the
KPZ universality class. Far from the origin, the accordance with the KPZ
conjecture is also observed, but the farther from the origin the slower the
convergence. 

\begin{figure}[ht]
 \centering
 \includegraphics[width=9cm]{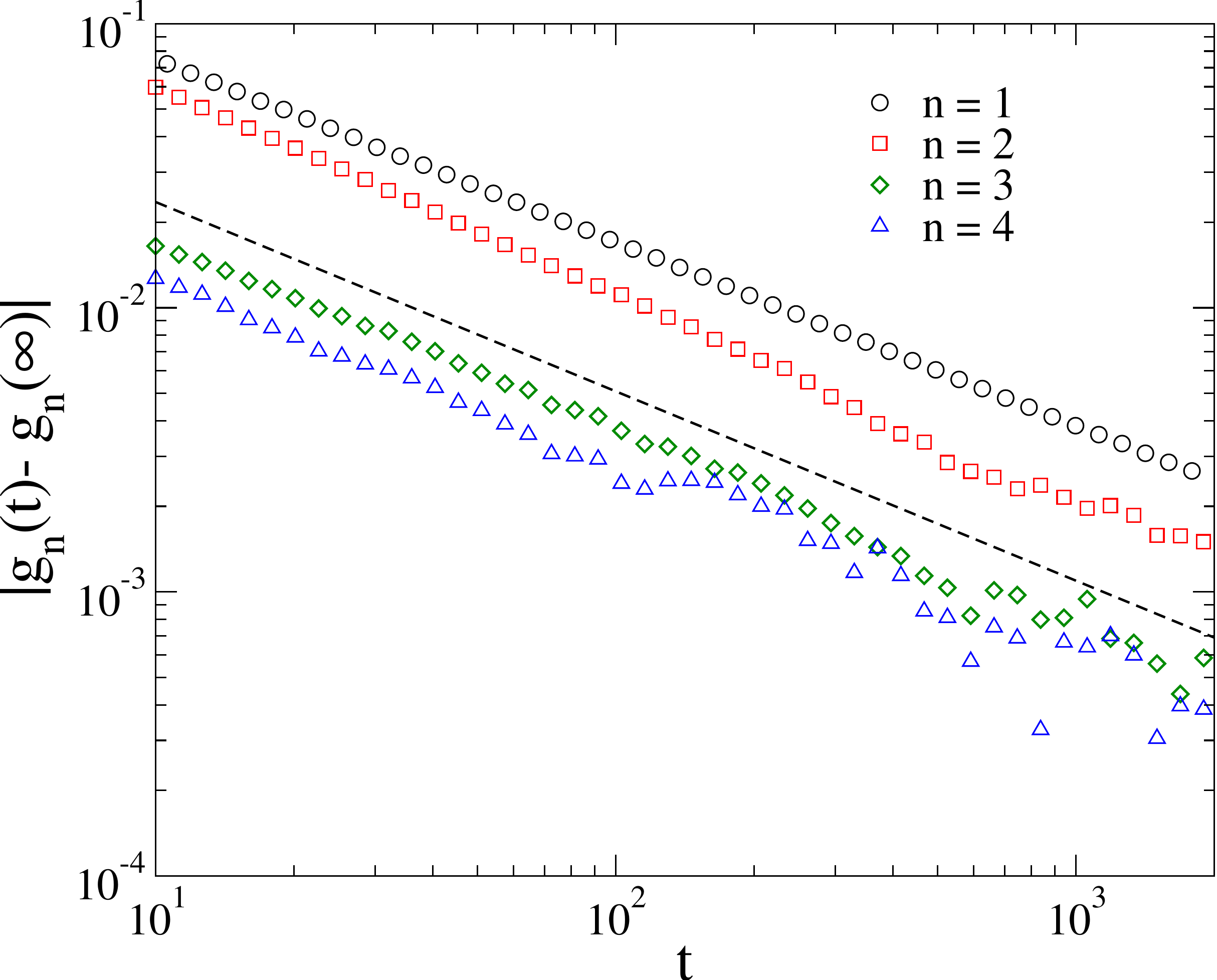}
 \caption{Time corrections in $g_n$ for WRSOS model at $j=0$. Dashed line
is a power law $t^{-2/3}$.}
 \label{fig:gnkk}
\end{figure}

The finite time corrections in the cumulants of WRSOS were also studied. The
shift in the mean approaches the GUE one with the correction $t^{-1/3}$
(figure~\ref{fig:shift1}). The quantities $g_n$, for $n = 1$, 3 and 4 have a
correction $t^{-2/3}$ while $g_2$ decays slightly faster than $t^{-2/3}$
(figure~\ref{fig:gnkk}) in analogy with the flat simulations of RSOS.

\section{Discussion}
\label{sec:conclu}

We have performed extensive simulations of models belonging to KPZ universality
class in order to probe the generality of the KPZ ansatz given by
equation~(\ref{eq:hpluscorr}) and to determine further corrections to this
equation. We have analyzed two classes of models. In the isotropic growth 
models, all points of the interface are used 
to perform averages, in contrast with the anisotropic ones, where the 
non-universal parameters vary along the surface, limiting the statistics to a single or 
a few points per sample.

During the numerical investigation, we have found out that the
determination of the non-universal parameter $\Gamma$ using the
asymptotic value of the scaled second cumulant $g_2=\lrangle{h^2}_c/t^{2/3}$, has
limited accuracy for some models, probably due to unknown puzzling corrections. We
propose an alternative way to obtain $\Gamma$ using the first cumulant
derivative by means of the quantity $g_1=s_\lambda
3(\lrangle{h}_t-v_\infty)t^{2/3}\rightarrow \Gamma^{1/3}\lrangle{\chi}$. It is
important to mention that noise in the numerical derivative counts against this
method. Here, we used very large statistics to smooth the curves. However, if
statistics is limited, as possibly in case of experiments, one can use numerical
procedures to smooth the derivative~\cite{NR}. We also considered scaled
cumulants $g_n=\lrangle{h^n}_c/(s_\lambda^n t^{n/3})$ of third and fourth order to
determine $\Gamma$. In general, we have observed that small deviations in $\Gamma$
may suggest a fake violation of the generalized KPZ ansatz. Estimates taken from 
$g_1$ give {very good agreement with the generalized KPZ 
conjecture for all investigated
models}. Dimensionless
universal cumulant ratios are not very sensitive to corrections in cumulants
as one can see in tables~\ref{tab:cumiso} and \ref{tab:uni},
 in which excellent agreements with GOE and GUE
distributions were found for flat and curved models, respectively. 

Comparing the non-universal parameters for the flat RSOS and BD models (table
\ref{tab:noniso}) with its droplet counterparts (table \ref{tab:nonaniso}), we see
that within the error bars they are equal, with exception of the shift
$\lrangle{\eta}$. On the other hand, in the Eden D model all
parameters are different for on- and off-lattice simulations. This is expected,
since the lattice constraint should change the strength of fluctuations.

In off-lattice simulations of the Eden model, we determined the shift in the height
distributions and our results support a crossover from $t^{-2/3}$ to
$t^{-1/3}$ in the first cumulant in relation to GUE, 
clarifying  an apparent exception to the correction $t^{-1/3}$ recently
suggested by Takeuchi~\cite{TakeuchiJstat}.

From the numerical side, we propose that the estimates of $\Gamma$ taken from
the first cumulant  are more reliable than those obtained from the higher order ones
since, in principle, we know little about the nature of terms beyond $\chi$ in
the KPZ ansatz. These terms may introduce anomalous corrections as, for
example, in the case of statistical dependence among random variables.
In fact, in contrast to the $t^{-2/3}$ correction  in $g_1$ obtained for all 
investigated models, we have not found a standard in the corrections of higher order
cumulants. 

The first cumulant derivative analysis for all 
investigated models are in agreement with an
additional term generalizing the KPZ ansatz to 
\begin{equation} 
h = v_\infty t+
s_\lambda(\Gamma t)^{1/3}\chi+\eta+\zeta t^{-1/3}+\ldots. 
\label{eq:ansatz2}
\end{equation}
Corrections $\Oc{t^{-1/3}}$ were reported also for some analytically tractable KPZ class
models~\cite{Ferrari}, suggesting that this term is a general 
property of KPZ systems. However, in the solution of the KPZ equation with a
sharp wedge initial condition a correction $\Oc{t^{-1}}$ was reported~\cite{SasaSpo1,SasaSpohnJsat}, 
possibly because the non-universal parameter $\zeta$ is zero in this case.

The generalized ansatz given by equation~(\ref{eq:ansatz2}) allows one to
speculate on the origin of the complicated corrections obtained for cumulants of
order $n\ge 2$. For instance, if $\eta$ and $\zeta$ are statistically dependent,
a term $t^{-1}$ will appear in the second cumulant in addition to  the leading
term $t^{-2/3}$. If the amplitude of $t^{-1}$ is much larger than that of
$t^{-2/3}$, one would observe  a slow crossover to $t^{-2/3}$. This scenario
explains the correction in $g_2$ observed for the RSOS model. Still in the field of
speculation, if $\chi$ and $\eta$ are statistically dependent the two first 
leading terms of the correction in $g_2$ would be $t^{-1/3}$ and $t^{-2/3}$.
Again, if the amplitude of the second leading term is much larger than the first
one, we could explain the anomalously slow exponent observed in the BD model as a crossover
between the two scaling regimes. However, a double power law regression, as that
used to determine the shift correction in Eden~D, does not support
such a conjecture.

\begin{figure}[b]
 \centering
 \includegraphics[width=8cm]{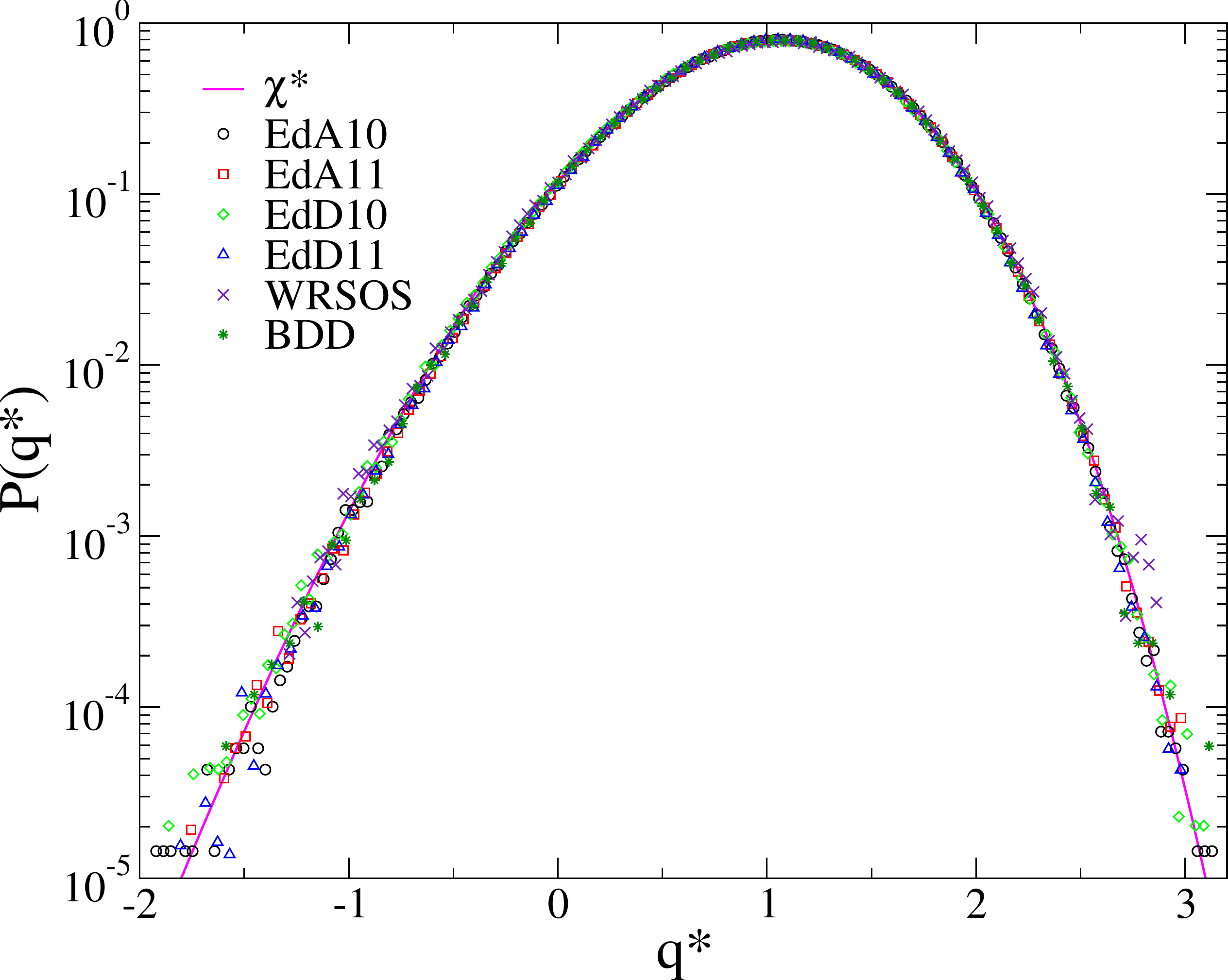} 

\includegraphics[width=8cm]{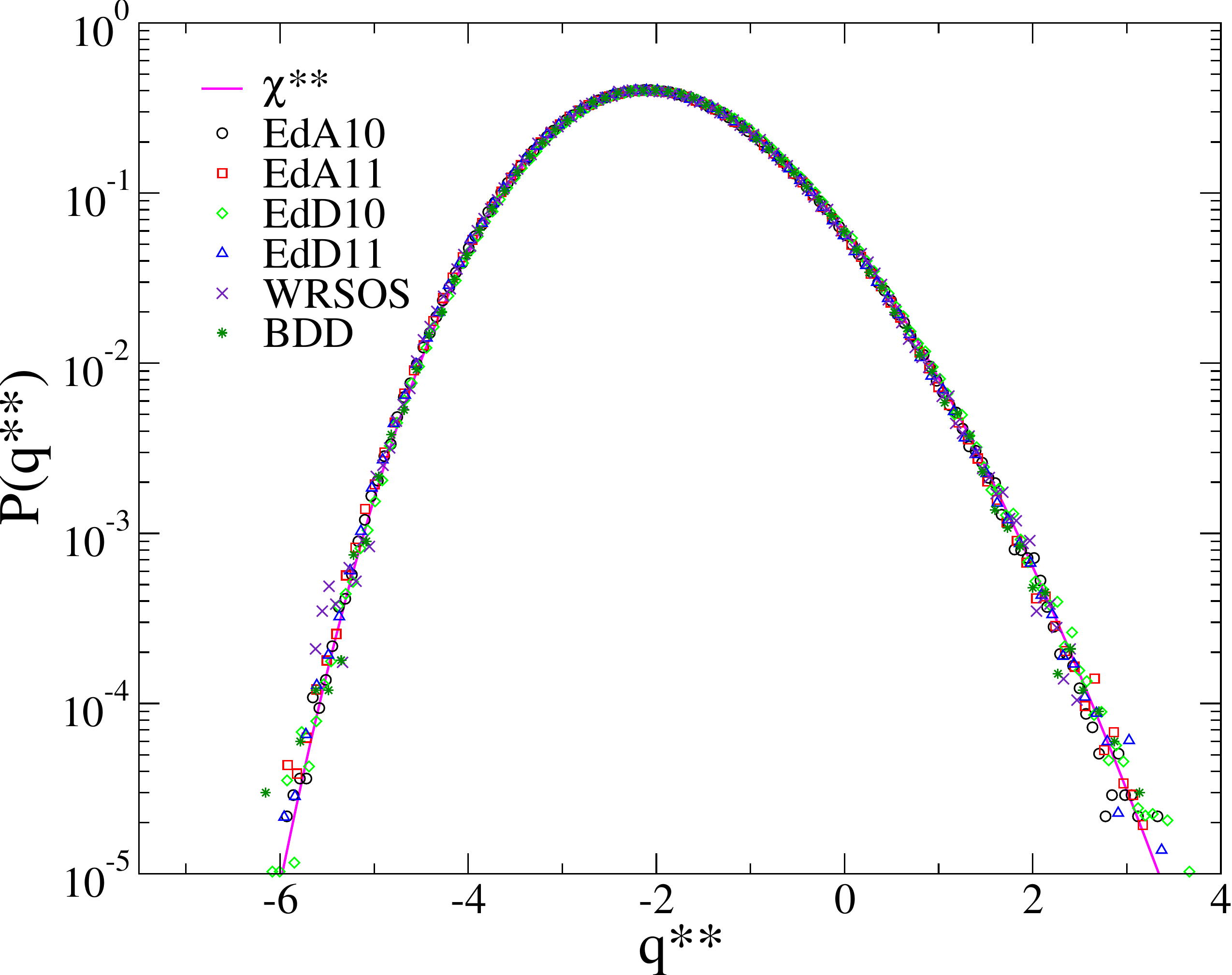}
 \caption{Rescaled distributions for all investigated models using
equations~(\ref{eq:qstar}) and (\ref{eq:qstar2}) in top and bottom panels,
respectively.}
 \label{fig:pofq}
\end{figure}

Now we discuss how universality in height distributions can be checked without 
an explicit knowledge of the parameter $\Gamma$.
Performing a rescaling using the  directly 
measurable parameters from tables \ref{tab:noniso}, 
\ref{tab:cumiso} and \ref{tab:nonaniso}, we can define the variable:
\[
q^* = \frac{h-v_\infty t-\lrangle{\eta}}{s_\lambda g_1t^{1/3}}.
\]
The scaled distributions for this variable, $P(q^*)$, agree remarkably well with the
GUE distribution scaled to unitary mean, $\chi^*=\chi/\lrangle{\chi}$, as shown in Figure~\ref{fig:pofq}, top.
Our entire analysis depends implicitly on $\lrangle{\chi}$, but does
not provide information about it beyond its negative sign, since $g_1<0$. We can
alternatively define a variable with asymptotic unitary variance using the
measurable parameter $g_2$ rather than $g_1$ to find
\begin{equation}
q^{**} = \frac{h-v_\infty t-\lrangle{\eta}}{g_2^{1/2}t^{1/3}} \rightarrow 
\frac{\chi}{\sqrt{\lrangle{\chi^2}_c}},
\label{eq:qstar2}
\end{equation}
with a mean depending on the dimensionless and measurable cumulant ratio $R$ given by
\[\lrangle{q^{**}}\equiv 
\mbox{sgn}(\lrangle{\chi}) R^{-1/2} = -1.9641.\]
Again, as shown in figure~\ref{fig:pofq} bottom, a very good collapse is obtained. We
remark that this kind of analysis could be very profitable to probe the
universality of height distributions in systems where they are not known
\textit{a priori}, such as high dimensional KPZ growth or models in universality
classes without analytical counterparts. Although the complete distribution are not
determined, once $\lrangle{\chi}$ or $\sqrt{\lrangle{\chi^2}_c}$  is fixed
everything concerning the HD can be derived.

\section*{Acknowledgments}
This work was partially supported by the Brazilian
agencies CNPq and FAPEMIG. We thank K. A. Takeuchi for the 
fruitful discussions on Ref.~\cite{TakeuchiJstat} and H. Spohn
for the discussions on Ref.~\cite{SasaSpo1}.

\section*{References}


\begin{thebibliography}{10}
\expandafter\ifx\csname url\endcsname\relax
  \def\url#1{{\tt #1}}\fi
\expandafter\ifx\csname urlprefix\endcsname\relax\def\urlprefix{URL }\fi
\providecommand{\eprint}[2][]{\url{#2}}

\bibitem{barabasi}
Barabasi A~L and Stanley H~E 1995 {\em Fractal Concepts in Surface Growth\/}
  (Cambridge, England: Cambridge University Press)

\bibitem{meakin}
Meakin P 1998 {\em Fractals, Scaling and Growth far from Equilibrium\/}
  (Cambridge, England: Cambridge University Press)

\bibitem{KPZ}
Kardar M, Parisi G and Zhang Y~C 1986 {\em Phys. Rev. Lett.\/} {\bf 56}
  889--892

\bibitem{Amir}
Amir G, Corwin I and Quastel J 2011 {\em Commun. Pure Appl. Math.\/} {\bf 64}
  466--537

\bibitem{Beijeren}
van Beijeren H 2012 {\em Phys. Rev. Lett.\/} {\bf 108}(18) 180601

\bibitem{TakeSano}
Takeuchi K~A and Sano M 2010 {\em Phys. Rev. Lett.\/} {\bf 104} 230601

\bibitem{TakeuchiSP}
Takeuchi K~A, Sano M, Sasamoto T and Spohn H 2011 {\em Sci. Rep.\/} {\bf 1} 34

\bibitem{TakeuchiJSP12}
Takeuchi K and Sano M 2012 {\em J. Stat. Phys.\/} {\bf 147}(5) 853--890

\bibitem{Yunker}
Yunker P~J, Lohr M~A, Still T, Borodin A, Durian D~J and Yodh A~G 2013 {\em
  Phys. Rev. Lett.\/} {\bf 110}(3) 035501

\bibitem{SasaSpo1}
Sasamoto T and Spohn H 2010 {\em Phys. Rev. Lett.\/} {\bf 104} 230602

\bibitem{Calabrese}
Calabrese P and Le~Doussal P 2011 {\em Phys. Rev. Lett.\/} {\bf 106}(25) 250603

\bibitem{SasaSpohnJsat}
Sasamoto T and Spohn H 2010 {\em J. Stat. Mech.: Theor. Exp.\/} {\bf 2010}
  P11013

\bibitem{Imamura}
Imamura T and Sasamoto T 2012 {\em Phys. Rev. Lett.\/} {\bf 108}(19) 190603

\bibitem{Doussal}
Doussal P~L and Calabrese P 2012 {\em J. Stat. Mech.\/} {\bf 2012} P06001

\bibitem{krugrev}
Kriecherbauer T and Krug J 2010 {\em J. Phys. A: Math. Theor.\/} {\bf 43}
  403001

\bibitem{johansson}
Johansson K 2000 {\em Commun. Math. Phys\/} {\bf 209} 437--476

\bibitem{PraSpo1}
Pr\"ahofer M and Spohn H 2000 {\em Phys. Rev. Lett.\/} {\bf 84} 4882--4885

\bibitem{PraSpo2}
Pr\"ahofer M and Spohn H 2000 {\em Physica A\/} {\bf 279} 342--352

\bibitem{TW1}
Tracy C and Widom H 1994 {\em Commun. Math. Phys.\/} {\bf 159} 151--174

\bibitem{SchehrEPL}
Rambeau J and Schehr G 2010 {\em Eur. Lett.)\/} {\bf 91} 60006

\bibitem{Alves11}
Alves S~G, Oliveira T~J and Ferreira S~C 2011 {\em Europhys. Lett.\/} {\bf 96}
  48003

\bibitem{Oliveira12}
Oliveira T~J, Ferreira S~C and Alves S~G 2012 {\em Phys. Rev. E\/} {\bf 85}
  010601

\bibitem{TakeuchiJstat}
Takeuchi K~A 2012 {\em J. Stat. Mech.\/} {\bf 2012} P05007

\bibitem{Ferrari}
Ferrari P and Frings R 2011 {\em J. Stat. Phys.\/} {\bf 144} 1--28

\bibitem{Kelling}
Kelling J and \'Odor G 2011 {\em Phys. Rev. E\/} {\bf 84} 061150

\bibitem{eden}
Eden M 1961 A two-dimensional growth process {\em Proceedings of Fourth
  Berkeley Symposium on Mathematics, Statistics, and Probability\/} vol~4 ed
  Neyman J (Berkeley,California: University of California Press) pp 223--239

\bibitem{Zabolitzky}
Zabolitzky J~G and Stauffer D 1986 {\em Phys. Rev. A\/} {\bf 34} 1523--1530

\bibitem{Batchelor99}
Batchelor M and Henry B 1991 {\em Phys. Lett. A\/} {\bf 157} 229--236

\bibitem{Alves06}
Alves S~G and Ferreira S~C 2006 {\em J. Phys. A: Math. Gen.\/} {\bf 39} 2843

\bibitem{Paiva07}
Paiva L~R and Ferreira S~C 2007 {\em J. Phys. A: Math. Theor.\/} {\bf 40} F43

\bibitem{Ferreira06}
Ferreira S~C and Alves S~G 2006 {\em J. Stat. Mech.: Theor. Exp.\/} {\bf 2006}
  P11007

\bibitem{BJP}
Alves S~G, Ferreira S~C and Martins M~L 2008 {\em Braz. J. Phys.\/} {\bf 38}
  81--86

\bibitem{Alves12}
Alves S~G and Ferreira S~C 2012 {\em J. Stat. Mech.\/} {\bf 2012} P10011

\bibitem{kk}
Kim J~M and Kosterlitz J~M 1989 {\em Phys. Rev. Lett.\/} {\bf 62}(19)
  2289--2292

\bibitem{krug}
Krug J 1997 {\em Adv. Phys.\/} {\bf 46} 139--282

\bibitem{Prolhac}
Prolhac S and Spohn H 2011 {\em Phys. Rev. E\/} {\bf 84} 011119

\bibitem{Bahman11}
Farnudi B and Vvedensky D~D 2011 {\em Phys. Rev. E\/} {\bf 83}(2) 020103

\bibitem{Robledo11}
Robledo A, Grabill C~N, Kuebler S~M, Dutta A, Heinrich H and Bhattacharya A 2011 {\em Phys. Rev. E\/} {\bf 83} 051604

\bibitem{NR}
Press W, Teukolsky S, Vetterling W and Flannery B 2007 {\em Numerical Recipes
  3rd Edition: The Art of Scientific Computing\/} (Cambridge University Press)
  ISBN 9780521880688

\end{thebibliography}

\providecommand{\newblock}{}

\end{document}